\documentclass[prl,aps,amssymb,reprint,twocolumn,superscriptaddress,longbibliography,10pt]{revtex4-2}
\usepackage{project_macros-public}
\usepackage{project_macros-private}
\usepackage{makecell}

\crefname{appendix}{Appendix}{Appendices}
\crefname{equation}{Eq.}{Eqs.}
\crefname{figure}{Fig.}{Figs.}
\crefname{table}{Table}{Tables}
\crefname{section}{Section}{Sections}
\creflabelformat{appendix}{[#2#1#3]}

\makeatletter
\AtBeginDocument{\let\LS@rot\@undefined}
\makeatother

\makeatletter
\renewcommand\onecolumngrid{
\do@columngrid{one}{\@ne}%
\def\set@footnotewidth{\onecolumngrid}
\def\footnoterule{\kern-6pt\hrule width 1.5in\kern6pt}%
}

\global\long\def\ket#1{\left| #1\right\rangle }%

\global\long\def\bra#1{\left\langle #1 \right|}%

\global\long\def\up{\uparrow}%

\global\long\def\down{\downarrow}%

\allowdisplaybreaks

\newif\ifarxiv
\newif\ifprl

\arxivtrue 

\begin{document}
\title{Ferromagnetism from the geometry of localized wavefunctions in moir\'e systems}

\paperAuthors

\author{Miguel Gonçalves}
\affiliation{Princeton Center for Theoretical Science, Princeton University, Princeton NJ 08544, USA}

\author{Sarang Gopalakrishnan}
\affiliation{Department of Electrical and Computer Engineering,
Princeton University, Princeton NJ 08544, USA}

\let\oldaddcontentsline\addcontentsline

\ifarxiv
  
\begin{abstract}
We present a mechanism for ferromagnetism in narrow bands consisting of Anderson-localized states. 
We exploit single-particle localization to derive a controlled theory of exchange interactions within the narrow band. 
For quasiperiodic systems with a half-filled moir\'e band, we show that the critical interaction strength for ferromagnetism is highly sensitive to the geometry of real-space overlaps between localized orbitals: we find well-defined ``resonances'' at which ferromagnetism sets in for interaction energies that are far lower than the gap to other bands. Near these resonances, all the approximations in our theory are controlled, so our critical point predictions are \emph{quantitative}.  
We show examples both in one and two dimensions. 
Our work identifies a route to ferromagnetism based on the geometry of real-space wavefunctions, distinct from previously found mechanisms based on the quantum geometry of Bloch bands.
\end{abstract}
\maketitle

Identifying the conditions under which electronic spins order ferromagnetically is a central question in many-body physics, and a surprisingly delicate one. Exact answers to this question are available only in a few isolated limits, and these answers are in some tension with one another. Lieb's theorem prohibits ferromagnetism at half-filling on bipartite lattices that are ``balanced'' (i.e., have the same number of $A$ and $B$ sublattice sites)~\cite{LiebMattis62,Lieb1989TwoTheorems,Miyao2018,PhysRevLett.126.100201}. On the other hand, the ground state of a half-filled exactly flat band is provably ferromagnetic~\cite{Mielke1991FerromagneticGroundStates,Mielke1991FurtherConsiderations, Mielke1992Kagome,Tasaki1992,MielkeTasaki1993,Mielke1999,PhysRevLett.99.026404,Katsura2010,Gulacsi2014,Gulacsi2014_phil}. The tension between these results is resolved by \emph{quantum geometry}~\cite{Kang2019StrongCoupling,Wu2020QuantumGeometryFerromagnetism,MielkeStauber2020,Bernevig2021TBGIII,Bernevig2021TBGV,PhysRevX.14.041004,Yu2025QGReview,zhang2025identifyinginstabilitiesquantumgeometry,chen2026quantumgeometricdipoletopologicalboost, Peotta2015,PhysRevLett.123.237002,PhysRevLett.124.167002,PhysRevB.101.060505,herzogarbeitman2022manybodysuperconductivitytopologicalflat,KukkaJonahRevisitingFlatbandSC2022,Torma2022,JonahSFWeightBounds2022,PhysRevLett.130.226001,han2025exactmodelschiralflatband}. In a model with a half-filled flat band (e.g., a kagome lattice), the microscopic lattice sites are not half-filled, so Lieb's theorem does not directly apply: because of the geometry of flat-band eigenstates, the flat band cannot be adiabatically ``detached'' from the other bands to form a freestanding lattice model. 
In realistic settings, bands are neither perfectly flat nor perfectly bipartite (as there is always some next-nearest-neighbor hopping), so the fate of ferromagnetism is unclear. 
The stability of ferromagnetism in narrow (but not perfectly flat) bands has been a topic of intense recent interest as such bands arise naturally in moir\'e materials~\cite{Balents2020StrongCorrelationsMoire,Andrei2021MarvelsMoire,Cao2018CorrelatedInsulator,Sharpe2019EmergentFerromagnetism,Chen2020CorrelatedChernFerromagnetism}: in these systems, interactions are large compared with the moiré bandwidth while remaining small compared with microscopic energy scales; moreover, there is clear evidence for magnetism. Yet, except for some special cases~\cite{TasakiNarrowBandFM_PRL94,TasakiNarrowBandFM_PRL95,Tasaki1996}, the stability of ferromagnetism away from the limit of perfectly flat bands has been addressed primarily using mean-field methods~\cite{Hu2025FerromagnetismVsAntiferromagnetism,espinosachampo2025magneticphasetransitionsprotected,oh2025magneticphasetransitionsdriven,kitamura2025quantumgeometricferromagnetismsingular}.

In the present work, we take a different route to establish the stability of ferromagnetism in narrow bands: we consider half-filled narrow \emph{localized} bands, which arise in random and quasiperiodic (e.g., moir\'e) systems. Specifically, we explore the Hubbard model with nearest- and next-nearest-neighbor hopping on quasiperiodic lattices, with the chemical potential tuned so that a narrow moir\'e band (containing a small fraction of the total lattice sites) is half-filled. The concrete models we consider are quasiperiodic Hubbard models in one or two dimensions, though our approach generalizes to other geometries (as well as to narrow Anderson-localized impurity bands beyond the quasiperiodic setting). It might seem at first sight that this approach would just complicate our analysis, since the interplay between localization and interactions is notoriously difficult to treat~\cite{efros2012electron, RevModPhys.66.261}. We show that, on the contrary, localization gives us a way of controllably treating interaction effects, using the localization length as a small parameter (see also Refs.~\cite{mahmood2021observation, PhysRevB.110.155137}). Our analysis proceeds as follows: we derive an effective Hamiltonian projected onto the narrow band, and exploit wavefunction localization to recast this as a hopping Hamiltonian for a single magnon. Using a combination of exact diagonalization and analytical techniques, we determine the stability of the fully polarized state against magnons: when it is stable, the ferromagnet is at least a local minimum of the energy, and we present numerical evidence that it is in fact the global minimum.

\begin{figure}[t]
    \centering
    \includegraphics[width=\columnwidth]{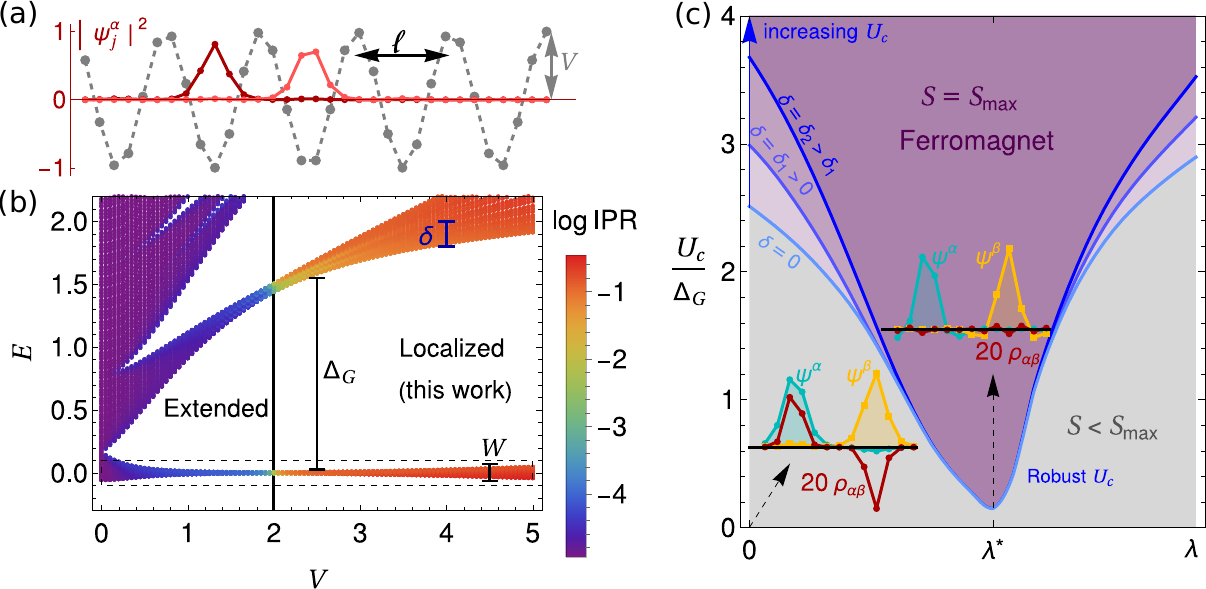}
    \caption{Problem and main results. (a) Illustration of a quasi-periodic potential (points in gray) forming a moiré pattern with length $\ell$. In the localized phase, the eigenstates in the lowest-energy narrow band are localized around the potential minima in each moiré cell. (b) Non-interacting spectrum of the 1D Aubry-André model, colored by the inverse participation ratio (IPR). We focus on half filling of the low-energy narrow band highlighted by the dashed box, within the localized phase. The bandwidth $W$ is much smaller than the gap $\Delta_G$ to higher-energy states, allowing a controlled narrow-band projection of the Hamiltonian when the relevant interaction scale satisfies $U\ll \Delta_G$. 
    (c) Critical interaction strength $U_c$ for the transition into the fully polarized ferromagnet as a function of next-nearest-neighbor hopping $\lambda$. Results are evaluated in a model that keeps higher-band states up to energy cutoff $\delta$. $\delta=0$ corresponds to projecting into the lowest-energy narrow-band only. For $\lambda \approx 0$, we find that $U_c$ depends on $\delta$ (so the single-band projection is not quantitatively accurate). However, for $\lambda$ near $\lambda^*$, $U_c$ is essentially independent of $\delta$; moreover, $U_c \ll \Delta_G$ so the single-band projection is fully controlled. We explain these observations in terms of the real-space overlap of neighboring moir\'e wavefunctions (illustrated in the figure): this overlap changes qualitatively from $\lambda = 0$ to $\lambda = \lambda^*$.}  
    \label{fig:1}
\end{figure}

Armed with this effective Hamiltonian, we estimate the threshold interaction strength $U_c$ at which ferromagnetism sets in. Dimensional analysis would suggest that this threshold should depend on some simple combination of the localization length and the width of the narrow band. 
In fact, we find that $U_c$ can change by over an order of magnitude even when such ``natural'' dimensional scales vary weakly. 
The key factor setting the strength of ferromagnetism is, instead, a \emph{geometrical} quantity, related to the real-space overlap between neighboring localized orbitals. Our results are illustrated in Fig.~\ref{fig:1}(c), where we show a
representative plot of $U_c$ as a function of a perturbation $\lambda$ that
breaks bipartiteness. When the spatial overlap profile $\rho_{\alpha\beta}$ is
resonant with the density profiles of the individual eigenstates $\alpha$ and
$\beta$, $U_c$ can substantially exceed the gap $\Delta_G$ to remote states.
This regime includes the bipartite limit $\lambda=0$, where we find that
including higher-energy states in the projection increases $U_c$~
\footnote{Although Lieb's theorem does not apply directly in our setup because
we are not at half-filling, in all bipartite cases we find
$U_c\gg\Delta_G$, consistent with the possibility that
$U_c\rightarrow\infty$. This is consistent with the lack of a clear topological obstruction to detaching the moir\'e band from the rest of the spectrum.}. By contrast, when the overlap profile becomes
off-resonant with the local density profiles, $U_c$ can become
significantly smaller than $\Delta_G$, making the band-projected theory
asymptotically controlled.

\paragraph{Models and Methods}

We study generalized Aubry-Andr\'e-Hubbard models in $d=1,2$ dimensions, given by the Hamiltonian
\begin{equation}
\begin{split}
H=&-\sum_{\mathbf{r},\eta,\{\mathbf{a}\}}t_{\mathbf{a}}c_{\eta,\mathbf{r}}^{\dagger}c_{\eta,\mathbf{r}+\mathbf{a}}+\textrm{h.c.}\\
&+\sum_{\mathbf{r},\eta}[V(\mathbf{r})-\mu]n_{\eta,\mathbf{r}}+U\sum_{\mathbf{r}}n_{\up,\mathbf{r}}n_{\down,\mathbf{r}},\\
V(\mathbf{r})=&V\sum_{j=1}^{d}\cos(2\pi\tau r_{j}+\phi_{j}),
\end{split}
\label{eq:general_AAH_model}
\end{equation}
where $n_{\eta,\mathbf r}=c_{\eta,\mathbf r}^{\dagger}c_{\eta,\mathbf r}$ and $\eta=\uparrow,\downarrow$. The displacement $\mathbf a$ includes first-, second-, and third-neighbor hoppings, with amplitudes $t=1$, $t_2$, and $t_3$, respectively. Unless otherwise stated, we choose
$\tau=(47+\sqrt{5})/58\simeq0.849$, which generates a moiré length
$\ell\simeq 1/(1-\tau)\simeq6.6$ in units of the atomic lattice spacing. We tune $\mu$ to half-fill the lowest-energy narrow band and use commensurate approximants $\tau=p/L$ to impose periodic boundary conditions, where $L$ is the linear number of sites, with $N=L^d$ the total number of sites. The approximants
used in the numerical calculations are listed in the Supplemental Material (SM). 

For the results shown in the main text, we project Eq.$\,$\ref{eq:general_AAH_model} onto the lowest-energy narrow band. In the basis of the $N_B$ narrow-band eigenstates, the projected Hamiltonian takes the form

\begin{equation}
\begin{aligned}\bar{H}= & \sum_{\alpha,\eta}\epsilon_{\eta,\alpha}\gamma_{\eta,\alpha}^{\dagger}\gamma_{\eta,\alpha}\\
 & +U\sum_{\alpha\alpha'\beta'\beta}V_{\up\down,\alpha\alpha'\beta'\beta}\gamma_{\up,\alpha}^{\dagger}\gamma_{\down,\alpha'}^{\dagger}\gamma_{\down,\beta'}\gamma_{\up,\beta} \, .
\end{aligned}
\label{eq:H_proj}
\end{equation}
with 
\begin{equation}
V_{\eta\eta',\alpha\alpha'\beta'\beta}=\sum_{{\bf r}}(\psi_{{\bf r}}^{\eta,\alpha})^{*}(\psi_{{\bf r}}^{\eta',\alpha'})^{*}\psi_{{\bf r}}^{\eta',\beta'}\psi_{{\bf r}}^{\eta,\beta}
\label{eq:projected_matrix_elements}
\end{equation}
Here $\gamma_{\eta,\alpha}^{\dagger}$ creates an electron with spin $\eta$ in the single-particle orbital $\alpha$ of the narrow band, with wave function $\gamma_{\eta,\alpha}^{\dagger}=\sum_{{\bf r}}\psi_{{\bf r}}^{\eta,\alpha}c_{\eta,\bf{r}}^{\dagger}$. For the $SU(2)$-symmetric models considered in the main text, the single-particle wave functions are spin independent, and the spin label on $\psi_{\mathbf r}^{\eta,\alpha}$ can be dropped. We nevertheless keep the spin label in the formal expressions, unless otherwise stated, since the same projected description applies straightforwardly to spin-dependent single-particle Hamiltonians.

In the localized phase, each of the narrow-band eigenstates is centered on a different moiré cell and we can therefore map the eigenstate index $\alpha$ to a moiré cell index, which will be useful to understand some of the results that follow \footnote{We note that this is no longer true if $\phi_j=0$, in which case the atomic limit ($V\rightarrow \infty$) eigenstates are exactly degenerate at mirror symmetric positions due to the existence of an exact inversion symmetry and will hybridize for any finite hopping. However, this degeneracy can be lifted by simply taking $\phi_j\neq0$. }.

\emph{Ferromagnetic ground state}.---We will specialize to the case where the lowest narrow band is at half filling. In the ferromagnetic phase, the ground state at this filling is a fully polarized Slater determinant, $\ket{\Uparrow}=\prod_{\alpha}\gamma_{\up,\alpha}^{\dagger}\ket 0$. The ferromagnet is unstable when adding ``simple'' excitations above it---in particular, spin-flip excitations, which correspond to removing a spin-$\uparrow$ particle and adding a spin-$\downarrow$ particle---decreases the energy. To study these excitations, it suffices to explore a two-body Hilbert space above the ferromagnetic ground state. This simplification allows us to reach large enough system sizes to converge the results to the thermodynamic limit.
The main text focuses on the lowest-energy narrow-band projection and on the local stability of ferromagnetism (i.e. one spin-flip with respect to $\ket{\Uparrow}$). To complement these results, in the SM we also show additional numerical results supporting the robustness of our conclusions upon including higher-energy states in the projection and upon considering additional spin sectors.

\emph{Charge excitations}.---Whether the ferromagnet is stable depends on the spin-flip spectrum; however, to understand the properties of spin-flip excitations (which are made up of a particle and a hole) it will be helpful to first discuss the even simpler problem of charge $\pm 1$ excitations above the fully polarized state. 
%
%
%
Charge $-1$ excitations consist of removing a spin-$\up$ electron from the fully polarized ground state $\ket{\Uparrow}$. The states $\gamma_{\up,\alpha}\ket{\Uparrow}$ are exact eigenstates of the Hamiltonian since they are annihilated by the interactions,
%
so the charge $-1$ spectrum is trivially $\epsilon^{(-)}_{\alpha}=-\epsilon_{\uparrow,\alpha}$.
On the other hand, charge $+1$ excitations 
are obtained by adding a single spin-$\down$ electron to the fully polarized state:
this added electron interacts with the spin-$\uparrow$ electrons, so its energy is dressed by Hartree shifts. 
%
The projected Hamiltonian closes exactly in the basis of states with exactly one spin-$\uparrow$ electron, and reduces to the one-body matrix
\begin{equation}
\begin{split}
R_{\alpha\beta}^{(+)}=&\bra{\Uparrow}\gamma_{\down,\alpha}\bar{H}\gamma_{\down,\beta}^{\dagger}\ket{\Uparrow}-E_{\Uparrow} \delta_{\alpha,\beta}\\
=&\delta_{\alpha\beta}\epsilon_{\down,\alpha}+U \bar{n}_{\up} \cdot \rho_{\alpha,\beta}^{\downarrow},
\end{split}
\label{eq:charge1}
\end{equation}
where   $[\bar{n}_{\up}]_{\bf r}=\sum_{\lambda}|\psi_{{\bf r}}^{\up,\lambda}|^{2}$ is the spin-$\up$ electron  density and we defined $[\rho_{\alpha\beta}^{\eta}]_{\bf r}=(\psi^{\eta,\alpha}_{\bf r})^{*}\psi^{\eta,\beta}_{\bf r}$. Here the dot product $a \cdot b$ is defined as $\sum\nolimits_{\mathbf{r}} a_{\mathbf{r}} b_{\mathbf{r}}$. The eigenvalues of $R_{\alpha\beta}^{(+)}$ give the exact charge +1 spectrum.
In the localized phase, the interacting contribution to Eq.$\,$\ref{eq:charge1} is dominated by its diagonal matrix elements $\alpha=\beta$, up to parametrically small corrections of the order of the overlaps between eigenstates localized at neighboring moiré cells. Further dropping the spin-dependence, the charge +1 spectrum is given by
\begin{equation}
    \epsilon^{(+)}_{\alpha} \approx \epsilon_{\alpha} + U\,\mathrm{IPR}(\psi^{\alpha}) \, .
    \label{eq:charge1_spectrum_simplified}
\end{equation}
Thus, to a good approximation, the charge +1 excitation eigenstates coincide with the single-particle eigenstates and are localized in the localized phase. 

\begin{figure}[t]
    \centering
    \includegraphics[width=\columnwidth]{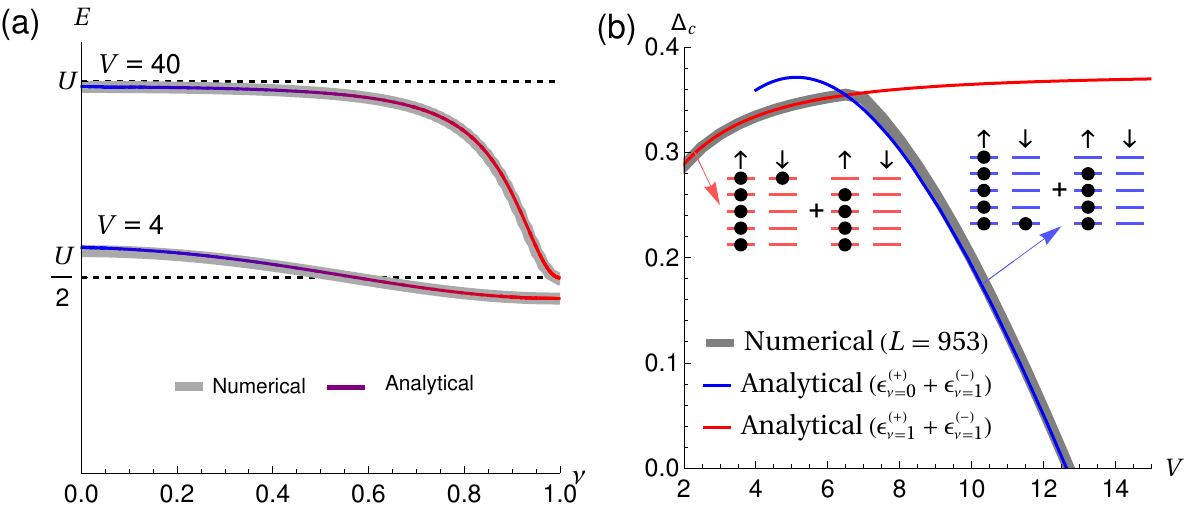}
    \caption{Numerical and analytical results for charge excitations for the 1D Aubry-André model with $t_2=t_3=0$. (a) Interacting contribution to the spectrum of charge excitations by diagonalization of the $U$-dependent term in Eq.$\,$\ref{eq:charge1}, as a function of the eigenstate label $\nu=\alpha/N_B$, where $\alpha=0,\cdots,N_B-1$. The analytical results using the interacting term in Eq.$\,$\ref{eq:charge1_spectrum_simplified} are also shown. (b) Charge gap $\Delta_c$ as a function of $V$ for $U=3/4$. The insets illustrate the dominant particle-hole excitations for small (red) and large (blue) $V$. The  analytical expressions are provided in the SM.}
    \label{fig:2}
\end{figure}

Although these eigenstates are simple, their energetics depends on $U$: the single-particle energy of an added spin-$\down$ particle is minimized when one puts it at the bottom of the narrow band. However, states at the bottom of the narrow band are strongly localized, so the interaction shift~\eqref{eq:charge1_spectrum_simplified} is largest for these states (and lowest for states near the top of the band, which are formed from bonding/anti-bonding combinations spread over two nearby sites). This competition explains the non-monotonic behavior of the charge gap $\Delta_c \equiv \textrm{min}_{\{\alpha,\beta \}}( \epsilon^{(+)}_{\alpha} + \epsilon^{(-)}_{\beta})$ shown in Fig.~\ref{fig:2}(b), which can also be captured analytically using the locator expansion (see SM). 

\paragraph{Magnons.---}We now turn to spin-flip excitations, which consist of adding one spin-$\down$ electron and removing a spin-$\up$ electron. 
These excitations are spanned by $\ket{\alpha,\beta}=\gamma_{\down,\alpha}^{\dagger}\gamma_{\up,\beta}\ket{\Uparrow}$ and can be obtained by diagonalizing the effective Hamiltonian (see SM),

\begin{equation}
\begin{aligned}
\mathcal{H}_{(\lambda\delta,\alpha\beta)}^{\textrm{\ensuremath{\down\up}}} = & 
 \bra{\lambda,\delta}\bar{H}\ket{\alpha,\beta}-E_{\Uparrow}\\
= & (R_{\lambda\alpha}^{(+)}-\epsilon_{\up,\beta}\delta_{\lambda\alpha})\delta_{\delta\beta}-UV_{\up\down,\beta\lambda\alpha\delta}
\end{aligned}
\label{eq:magnon_H_full}
\end{equation}
Here $R_{\lambda\alpha}^{(+)}$ is the charge +1 matrix defined in
Eq.~\eqref{eq:charge1} and the last term is the binding matrix, defined in terms of the projected matrix elements in Eq.$\,$\ref{eq:projected_matrix_elements}.

Fig.$\,$\ref{fig:3}(a) shows representative results for the spin-flip spectrum obtained by diagonalizing the Hamiltonian in Eq.$\,$\ref{eq:magnon_H_full}. Recall that the single-particle states are localized. At small $U$, the lowest spin-flip excitation consists of removing a spin-$\up$ particle from the highest-energy single-particle state, and adding a spin-$\down$ particle to the lowest (dressed) single-particle state. This is an ``unbound'' particle-hole excitation [illustrated in blue in Fig.~\ref{fig:3}(a)], which is dominant when $U \ll W$. In this regime, the ferromagnet is unstable since one can decrease the (dominant) kinetic energy by flipping spins and occupying lower-lying orbitals. 

As $U$ increases past $W$ [specifically, at $U=U^*\approx W/\textrm{IPR}(\psi^{0})$], the interaction energy cost of the added particle dominates over the kinetic energy gain. The lowest-lying particle-hole excitation switches from being an unbound particle-hole pair, to a spin-$\down$ particle created in precisely the same orbital as the vacated spin-$\up$ particle: namely, a spin-flip or magnon. In localized systems, this binding transition is a first-order \emph{spectral transition}, since the magnon state is macroscopically distinct from the unbound particle-hole state.
%
%
Importantly, however, the onset of magnons does not correspond to the transition point into the fully polarized ferromagnet.
Rather, the collective magnon
modes become the lowest spin-flip modes, while still lying below the fully polarized state, as shown in the inset of Fig.~\ref{fig:3}(a). The fully polarized state becomes locally stable only at a larger interaction $U_c>U^*$, where the lowest magnon energy crosses zero. We note that although this criterion only tests stability against one-spin-flip excitations, in the SM we show exact diagonalization results in all spin sectors, consistent with $\ket{\Uparrow}$ being the global ground state for $U>U_c$. 

We now develop an analytical description of the collective magnon modes and of the key ingredients controlling $U_c$. We focus on the localized, large-$V$ regime for $U>U^*$, where the magnon modes are separated from the particle-hole continuum. In this regime, the low-energy magnon subspace is spanned by
$\ket{\alpha,\alpha}$, where the spin-$\uparrow$ hole and spin-$\downarrow$ particle form a bound pair occupying the same localized moiré orbital. The only energy cost of these states comes from overlaps between orbitals centered on different moiré cells, which vanish as $V\rightarrow\infty$. In the $V \to \infty$ limit, the
$\ket{\alpha,\alpha}$ states become zero-energy eigenstates. At finite $V$ in the localized phase, the magnons overlap, and interactions hybridize magnons on different orbitals.


\begin{figure}[bt]
    \centering
    \includegraphics[width=\columnwidth]{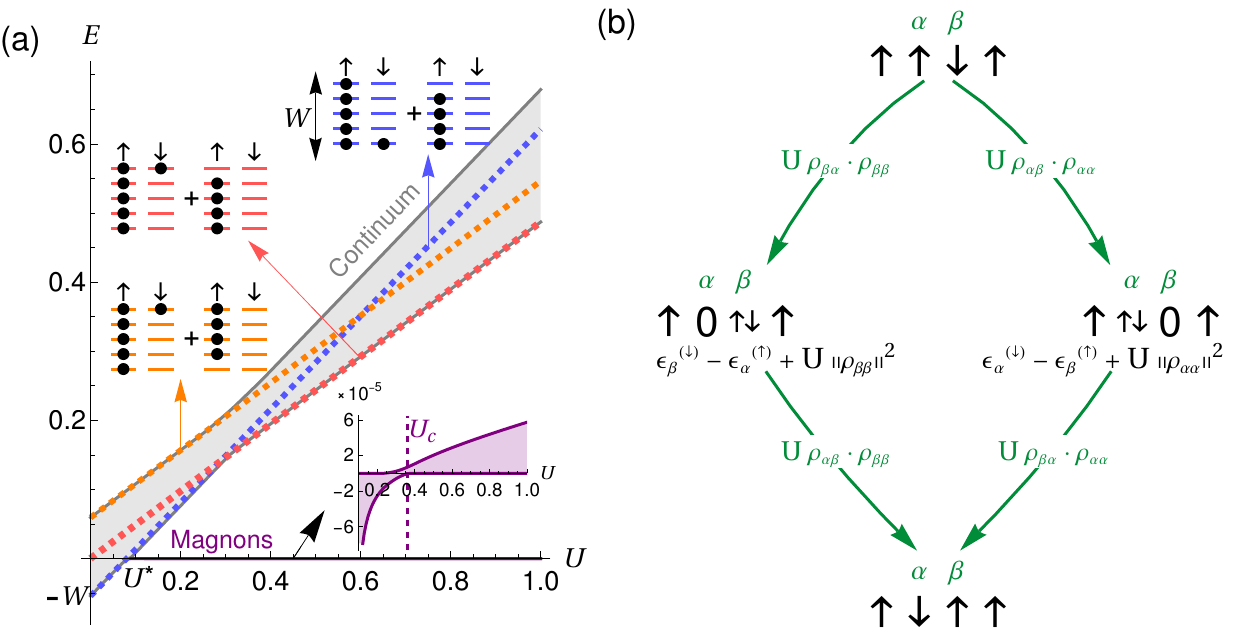}
    \caption{Spin-flip excitations and magnons. (a) Example magnon spectrum for $V=3.5$, $t_2=-0.7$ and $L=225$. The upper insets illustrate different competing excitations in the particle-hole continuum. The lower inset shows a close-up of the narrow magnon bound-state branch. For $U>U^*$ the collective magnon modes become the lowest-energy spin-flip excitations, while for $U>U_c$ the lowest magnon-energy crosses zero and the fully polarized ferromagnet becomes the ground state. (b) Illustration of the virtual magnon superexchange processes entering Eq.$\,$\ref{eq:magnon_main_text}. 
    }
    \label{fig:3}
\end{figure}

Projecting $\bar{H}$ into the low-energy subspace spanned by $\{ |\alpha, \alpha\rangle \}$ and integrating out virtual high-energy processes involving basis elements  ${\ket{\alpha,\beta}}$ with $\alpha \neq \beta$, we get that the magnon spectrum coincides with that of an effective Heisenberg spin Hamiltonian (see SM),

\begin{equation}
\begin{aligned}
H_{S}
&=
\sum_{\alpha<\beta}
J_{\alpha\beta}
\left(
\mathbf S_{\alpha}\cdot\mathbf S_{\beta}
-\frac{1}{4}
\right), \\
\mathbf S_{\alpha}
&=
\frac{1}{2}
\begin{pmatrix}
\gamma_{\uparrow,\alpha}^{\dagger} &
\gamma_{\downarrow,\alpha}^{\dagger}
\end{pmatrix}
\boldsymbol{\sigma}
\begin{pmatrix}
\gamma_{\uparrow,\alpha} \\
\gamma_{\downarrow,\alpha}
\end{pmatrix},
\end{aligned}
\label{eq:Heff_Heisenberg}
\end{equation}
where $\boldsymbol{\sigma}$ is a vector of Pauli matrices, and the exchange couplings are given by 
\begin{equation}
J_{\alpha\beta}\approx-2U\|\rho_{\alpha\beta}\|^{2}\Bigg[1-\frac{\cos^{2}\theta_{\alpha\beta}}{1+\frac{\Delta_{\alpha\beta}}{U\|\rho_{\alpha\alpha}\|^{2}}}-\frac{\cos^{2}\theta_{\beta\alpha}}{1+\frac{\Delta_{\beta\alpha}}{U\|\rho_{\beta\beta}\|^{2}}}\Bigg] \, ,
\label{eq:magnon_main_text}
\end{equation}
where
$\Delta_{\alpha\beta}\equiv\epsilon_{\alpha}-\epsilon_{\beta}$,  $\cos\theta_{\alpha\beta}=(\rho_{\alpha\alpha}\cdot\rho_{\alpha\beta})/(\|\rho_{\alpha\alpha}\|\|\rho_{\alpha\beta}\|)$, $[\rho_{\alpha\beta}]_{\bf r}$ was defined below Eq.$\,$\ref{eq:charge1} (we drop the spin indices from this point on for simplicity), and we defined the ``norm'' $\Vert a \Vert \equiv \sqrt{a \cdot a}$. Note that $\|\rho_{\alpha\alpha}\|^{2}\equiv\textrm{IPR}(\psi^\alpha)$. Because $\|\rho_{\alpha\beta}\|$ decays exponentially with the separation between orbitals $\alpha$ and $\beta$, these can be restricted to lie in nearest-neighbor moiré cells.

In the limit $U \times \mathrm{IPR}(\psi^\alpha) \gg W$ for all states in the band, Eq.~\eqref{eq:magnon_main_text} can be further simplified: the term in brackets is approximately $1 - \{\cos^2(\theta_{\alpha\beta}) + \cos^2(\theta_{\beta\alpha})\}$. The ``1'' is simply the ferromagnetic exchange interaction between neighboring localized orbitals, i.e., the direct energy cost of occupying opposite-spin orbitals in neighboring moiré cells $\alpha$ and $\beta$. The other terms are antiferromagnetic, and represent superexchange-like processes in which the spin-$\down$ particle virtually dissociates from its spin-$\up$ partner: these terms arise because the orbital $\ket{\alpha}$ is not a true charge eigenstate [cf. Eq.~\eqref{eq:charge1}].  
%
%
Although these virtual processes involve a superexchange-like mechanism, they differ from conventional superexchange in an important way, which we will see to have important physical consequences. 
Note that since we are working in the narrow-band
eigenbasis, the virtual high-energy subspace can only be accessed through the projected interaction matrix elements rather than through hopping. Moreover, these virtual corrections can be of the same order as the direct exchange, as
made explicit by Eq.~\eqref{eq:magnon_main_text}. The approximation is, nonetheless, controlled in the localized phase because higher-order virtual processes involve higher powers of the small inter-orbital overlap $\|\rho_{\alpha\beta}\|$ (see
the SM).

\paragraph{Mechanism for ferromagnetism.---}
From the analytical expression in Eq.~\eqref{eq:magnon_main_text} the mechanism controlling the onset of ferromagnetism can be made transparent. In the flat-band limit, $\Delta_{\alpha\beta}=0$, the exchange coupling reduces to (see SM)

\begin{figure}[bt]
    \centering
    \includegraphics[width=\columnwidth]{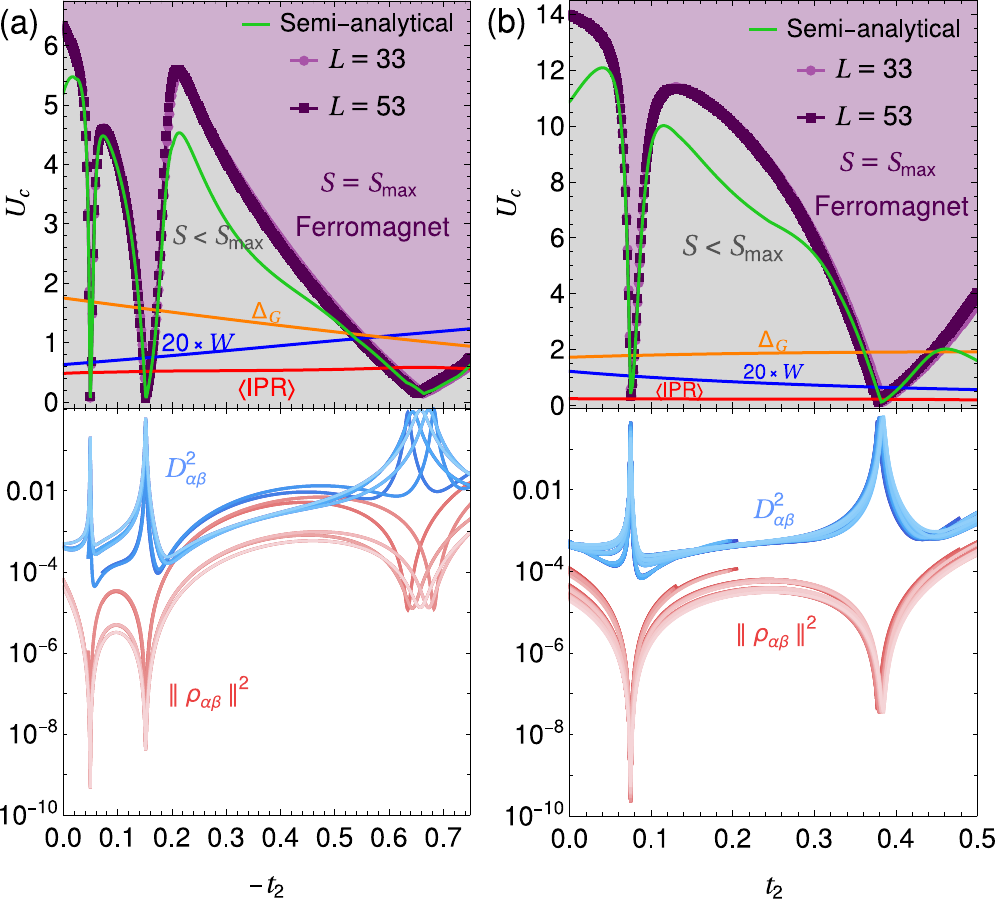}
    \caption{Ferromagnets induced by wavefunction detuning.
    Top panels: $U_c$ as a function of $t_2$ for different system sizes. Bottom panels: squared overlap norm $\|\rho_{\alpha\beta}\|^2$ and wavefunction detuning $\mathcal D_{\alpha\beta}^2$ for the dominant nearest-neighbor bonds entering the estimate of $U_c$ in Eq.~\eqref{eq:Uc} (shown in green). We restrict to pairs with $\Delta_{\alpha\beta}>\delta_r=0.01$, for which the off-diagonal particle-hole states $\ket{\alpha,\beta}$ can be reliably integrated out.  We also show the single-particle gap $\Delta_G$, the narrow-band bandwidth $W$, and the average inverse participation ratio $\langle \mathrm{IPR}\rangle$ of the narrow-band eigenstates. (a) Results for the 1D model with $t_3=0$. (b) Results for the 2D model with $t_2=-1.25t_3$.   }
    \label{fig:4}
\end{figure}

\begin{equation}
J_{\alpha\beta}\approx-2U\|\rho_{\alpha\beta}\|^{2}\Big(\mathcal{D}_{\alpha\beta}^{2}+\mathcal{O}(\Delta_{\alpha\beta}/U)\Big),
\label{eq:J_W=0}
\end{equation}
where we defined the \textit{wavefunction detuning}
\begin{equation}
\mathcal D_{\alpha\beta}
=
\sqrt{1-\cos^{2}\theta_{\alpha\beta}
-\cos^{2}\theta_{\beta\alpha}}
\approx
\left\|
\left(
1-P_{\{\rho_{\alpha\alpha},\rho_{\beta\beta}\}}
\right)
\hat\rho_{\alpha\beta}
\right\|.
\label{eq:Dab_def}
\end{equation}
where the last approximation is valid in the localized phase (see SM).
$P_{\{\rho_{\alpha\alpha},\rho_{\beta\beta}\}}$ denotes the projector onto the subspace spanned by the two real-space density profiles $\rho_{\alpha\alpha}$ and $\rho_{\beta\beta}$, and $\hat\rho_{\alpha\beta}
\equiv\rho_{\alpha\beta}/\|\rho_{\alpha\beta}\|$.
Eq.$\,$\ref{eq:J_W=0} implies that $J_{\alpha\beta}<0$ in the flat-band limit and therefore that the Hamiltonian in Eq.$\,$\ref{eq:Heff_Heisenberg} is positive-semidefinite (which can be easily shown by noting that $1/4-\mathbf S_{\alpha}\cdot \mathbf S_{\beta}$ is a projector into the two-site singlet subspace). As a consequence, the (zero-energy) ferromagnetic ground-state is necessarily stable to magnon excitations.
Equation \ref{eq:J_W=0} also shows that the flat-band ferromagnetic exchange coupling factorizes into two ingredients. The overlap norm $\|\rho_{\alpha\beta}\|^{2}$ sets the absolute magnitude of the coupling, while $\mathcal D_{\alpha\beta}^{2}$ measures the fraction of this overlap-induced ferromagnetic exchange that survives cancellation from the antiferromagnetic virtual processes. Geometrically, the wavefunction detuning $\mathcal D_{\alpha\beta}$ measures how much the overlap profile $\hat\rho_{\alpha\beta}$ is misaligned with the subspace spanned by the density profiles of eigenstates $\alpha$ and $\beta$. The larger this detuning is, the larger the finite-bandwidth correction controlled by $\Delta_{\alpha\beta}/U$ must be to destabilize the large-$U$ ferromagnet, by turning the exchange couplings antiferromagnetic.

The critical interaction can be estimated more explicitly by inspecting when the dominant exchange couplings change sign. Choosing the bond orientation such that $\Delta_{\alpha\beta}>0$, we obtain

\begin{equation}
\tilde{U}_{c}=\underset{\Delta_{\alpha\beta}>\delta_r}{\underset{\{\alpha,\beta\}\in\textrm{NN}}{\textrm{max}}}\frac{2\Delta_{\alpha\beta}/(\|\rho_{\alpha\alpha}\|\|\rho_{\beta\beta}\|)}{\Lambda_{\alpha\beta}+\sqrt{\Lambda_{\alpha\beta}^{2}+4\mathcal{D}_{\alpha\beta}^{2}}}\,,
\label{eq:Uc}
\end{equation}
where we defined the anti-symmetric tensor $\Lambda_{\alpha\beta}=(\|\rho_{\beta\beta}\|/\|\rho_{\alpha\alpha}\|)(1-\cos^{2}\theta_{\beta\alpha})-(\alpha\leftrightarrow\beta)$ and "NN" denotes orbitals at nearest-neighbor moiré cells. Note that we use a finite cutoff $\delta_r$ to exclude nearly energy-resonant pairs, for which the off-diagonal states $\ket{\alpha,\beta}$ with $\alpha\neq\beta$ cannot be reliably integrated out and must be retained explicitly.

Figures~\ref{fig:4}(a,b) show examples in one and two dimensions where ferromagnetism is stabilized at critical interactions much smaller than the gap $\Delta_G$ to remote bands. The minima of $U_c$ occur near points where $\mathcal D_{\alpha\beta}$ sharply increases, as shown in Fig.~\ref{fig:4}. By contrast, the bandwidth and the average IPR of the narrow band eigenstates vary smoothly across the same parameter range. This demonstrates that the sharp reduction of $U_c$ is not driven by a conventional bandwidth or localization effect, but by the wavefunction detuning here introduced.

\paragraph{Discussion.---}

We identified a mechanism for ferromagnetism in narrow bands of Anderson-localized states, governed by the geometry of single-particle eigenstate overlaps. In particular, we showed that the critical interaction strength required to stabilize a fully polarized ferromagnet can be reduced well below the gap to higher-energy bands by increasing the wavefunction detuning between neighboring localized eigenstates, and thus suppressing higher-order processes where the ``particle'' and ``hole'' that make up a magnon get far away from one another. 
The wavefunction detuning here introduced is distinct from the more conventional quantum geometry, which corresponds to Kohn's localization tensor in the absence of translational invariance \cite{Yu2025QGReview}. While quantum geometry characterizes the spatial extent of single-particle states, the wavefunction detuning here introduced probes the relative geometry of pairs of localized eigenstates. In particular, it measures the component of the real-space overlap profile that remains effective in the ferromagnetic exchange. In this sense, it captures a real-space geometric property of eigenstate overlaps that is particularly natural in Anderson localized phases.

The framework introduced here for studying magnetism in localized bands has wide applications beyond the current setting. We explicitly derived the effective model for magnons for quasiperiodic systems, but our derivation generalizes directly to disordered impurity bands. It would be interesting to explore magnon spectra (as well as the localization of finite-frequency magnons) in such bands more generally. 
More broadly, our work opens an important direction for the study of magnetism in correlated systems without translation symmetry, where the interplay between single-particle localization, multifractality, and interactions can generate
mechanisms not available in periodic Bloch bands. We expect this landscape to be further enriched in systems with nontrivial topology, where the real-space geometric effects identified here may coexist with, or be modified by, Berry curvature, quantum metric, and topological obstructions. 

\paragraph{Acknowledgements}
We thank Jonah Herzog-Arbeitman, Shi-Zeng Lin, Daniel Muñoz-Segovia, Pok Man Tam, and Carlo Vanoni for fruitful discussions.
\fi

\renewcommand{\addcontentsline}[1]{}
\nocite{REVTEX42Control,apsrev42Control}
\bibliographystyle{apsrev4-2}
\bibliography{revtex-control,refs}
\let\addcontentsline\oldaddcontentsline

\renewcommand{\thetable}{S\arabic{table}}
\renewcommand{\thefigure}{S\arabic{figure}}
\renewcommand{\theequation}{S\arabic{section}.\arabic{equation}}
\onecolumngrid
\pagebreak
\thispagestyle{empty}

\cleardoublepage

\begin{center}
	\textbf{\large Supplementary Information for ``\titlePaper{}"}\\[.2cm]
\end{center}

\appendix
\setcounter{secnumdepth}{3} 
\renewcommand{\thesection}{\Roman{section}}
\tableofcontents
\let\oldaddcontentsline\addcontentsline
\newpage
\section{Details on numerical calculations and commensurate approximants}

In all numerical calculations we used commensurate approximants of the irrational wave vector $\tau=\frac{47+\sqrt{5}}{58}\approx 0.849$ (see Eq.$\,$\ref{eq:general_AAH_model} of the main text for definition). This method allows us to impose periodic boundary conditions without any edge defect and therefore avoid probing edge physics. We write these approximants as $\tau=\tau_{c}=p/L$, where $L$ is the number of sites of the system in each direction. 

We list the sequence of some commensurate approximants in Table~\ref{tab:commensurate_approximants}. In this table, we also report the number of states in the narrow-band for each approximant in one dimension, $N_{B}^{1D}$, which follows the Fibonacci sequence and is approximately $15\%$ of the total number of single-particle states. In two dimensions, the number of states in the narrow-band is simply given by $N_{B}^{2D}=(N_{B}^{1D})^2$. 

\begin{table}[h]
\centering
\begin{tabular}{r r r}
\hline
$p$ & $L$ & $N_{B}^{1D}$\\
\hline
17   & 20    & 3\\
28   & 33    & 5\\
45   & 53    & 8\\
73   & 86    & 13\\
118  & 139   & 21\\
191  & 225   & 34\\
309  & 364   & 55\\
500  & 589   & 89\\
809  & 953   & 144\\
1309 & 1542  & 233\\
2118 & 2495  & 377\\
3427 & 4037  & 610\\
5545 & 6532  & 987\\
8972 & 10569 & 1597\\
\hline
\end{tabular}
\caption{Some commensurate approximants $\tau_{c}=p/L$ of $\tau=(47+\sqrt{5})/58$ with the corresponding number of states in the narrow-band in 1D, $N_{B}^{1D}$.  In 2D, the total number of states in the narrow-band is  $N_{B}^{2D}=(N_{B}^{1D})^2$, while the total number of sites is $L^2$.}
\label{tab:commensurate_approximants}
\end{table}

\section{Charge excitations and locator expansion}

In this section, we provide details on the charge +1 excitations. We start by deriving the effective one-body Hamiltonian for these excitations in  Sec.$\,$\ref{sec:charge1_Heff}. We then derive a locator expansion in Sec.$\,$\ref{sec:charge1_locator_expansion} around $V=\infty$ to obtain analytical expressions for the charge +1 excitation spectrum, that we show in detail in Sec.$\,$\ref{sec:charge1_analytical}.
For the locator expansion, we will focus on the one-dimensional Aubry-Andr\'e model with $t_{2}=t_{3}=0$ for simplicity, but it can be straightforwardly generalized to the remaining models studied in the main text.

\subsection{Effective charge +1 Hamiltonian}

\label{sec:charge1_Heff}

Charge +1 excitations above the fully polarized state $\ket{\Uparrow}=\prod_{\alpha}\gamma_{\up,\alpha}^{\dagger}\ket 0$ are spanned by the basis elements $\gamma_{\down,\alpha}^{\dagger}\ket{\Uparrow}$. To derive the effective one-body Hamiltonian, we evaluate separately the projected non-interacting and interacting contributions. Writing

\begin{equation}
    \bar{H} = \bar{H}_0 + \bar{H}_U
\end{equation}
with
\begin{equation}
\bar{H}_{0}=\sum_{m,\eta}\epsilon_{m}\gamma_{\eta,m}^{\dagger}\gamma_{\eta,m},
\qquad
\bar{H}_{U}=U\sum_{mm'n'n}V_{\up\down,mm'n'n}\gamma_{\up,m}^{\dagger}\gamma_{\down,m'}^{\dagger}\gamma_{\down,n'}\gamma_{\up,n},
\end{equation}
we will evaluate the single-particle excitation matrix defined in the main text, that we reproduce here for clarity,
\begin{equation}
R_{\alpha\beta}^{(+)}=\bra{\Uparrow}\gamma_{\down,\alpha}\bar{H}\gamma_{\down,\beta}^{\dagger}\ket{\Uparrow}-E_{\Uparrow}\delta_{\alpha\beta}.
\end{equation}
For the non-interacting part, since $\bar{H}_{0}\ket{\Uparrow}=E_{\Uparrow}\ket{\Uparrow}$ and $[\bar{H}_{0},\gamma_{\down,\beta}^{\dagger}]=\epsilon_{\beta}\gamma_{\down,\beta}^{\dagger}$, we obtain
\begin{equation}
\bra{\Uparrow}\gamma_{\down,\alpha}\bar{H}_{0}\gamma_{\down,\beta}^{\dagger}\ket{\Uparrow}=(E_{\Uparrow}+\epsilon_{\beta})\delta_{\alpha\beta}.
\end{equation}
For the interaction term, $\bar{H}_{U}\ket{\Uparrow}=0$, so we have:
\begin{equation}
\bra{\Uparrow}\gamma_{\down,\alpha}\bar{H}_{U}\gamma_{\down,\beta}^{\dagger}\ket{\Uparrow}=\bra{\Uparrow}\gamma_{\down,\alpha}[\bar{H}_{U},\gamma_{\down,\beta}^{\dagger}]\ket{\Uparrow}.
\end{equation}
Further using $\{\gamma_{\down,n'},\gamma_{\down,\beta}^{\dagger}\}=\delta_{n'\beta}$ and $\{\gamma_{\up,n},\gamma_{\down,\beta}^{\dagger}\}=0$, we have
\begin{equation}
[\gamma_{\up,m}^{\dagger}\gamma_{\down,m'}^{\dagger}\gamma_{\down,n'}\gamma_{\up,n},\gamma_{\down,\beta}^{\dagger}]
=-\delta_{\beta n'}\gamma_{\up,m}^{\dagger}\gamma_{\down,m'}^{\dagger}\gamma_{\up,n}.
\end{equation}
Acting with $\gamma_{\down,\alpha}$ from the left and then taking the expectation value in $\ket{\Uparrow}$ gives
\begin{equation}
\bra{\Uparrow}\gamma_{\down,\alpha}[\gamma_{\up,m}^{\dagger}\gamma_{\down,m'}^{\dagger}\gamma_{\down,n'}\gamma_{\up,n},\gamma_{\down,\beta}^{\dagger}]\ket{\Uparrow}
=\delta_{\alpha m'}\delta_{\beta n'}\bra{\Uparrow}\gamma_{\up,m}^{\dagger}\gamma_{\up,n}\ket{\Uparrow} = \delta_{\alpha m'}\delta_{\beta n'} \delta_{mn}.
\end{equation}
which yields
\begin{equation}
\bra{\Uparrow}\gamma_{\down,\alpha}\bar{H}_{U}\gamma_{\down,\beta}^{\dagger}\ket{\Uparrow}=U\sum_{m}V_{\up\down,m\alpha\beta m}.
\label{eq:R_interaction_derivation_SM}
\end{equation}

As discussed in the main text, the projected many-body problem closes exactly in this sector and reduces to the one-body matrix
\begin{equation}
R_{\alpha\beta}^{(+)}=\delta_{\alpha\beta}\epsilon_{\alpha}+R^{U}_{\alpha\beta},
\qquad
R^{U}_{\alpha\beta}=U \bar{n}_{\up} \cdot \rho_{\alpha,\beta}^{\downarrow},
\label{eq:charge1_SM_exact}
\end{equation}
where
\begin{equation}
 [\bar n_{\eta}]_{\bf r}=\sum_{\lambda}|\psi_{\bf r}^{\eta,\lambda}|^{2}
\end{equation}
is the spin-$\eta$ electron density, and $[\rho_{\alpha\beta}^{\eta}]_{\bf r}=(\psi^{\eta,\alpha}_{\bf r})^{*}\psi^{\eta,\beta}_{\bf r}$. Additionally, recall that the dot product $a \cdot b$ is defined as $\sum\nolimits_{\mathbf{r}} a_{\mathbf{r}} b_{\mathbf{r}}$. The eigenvalues of $R_{\alpha\beta}^{(+)}$ give the exact charge +1 spectrum in the band-projected theory.

Deep in the localized phase, each narrow-band eigenstate is peaked on a different moir\'e cell. Therefore $(\psi_{\bf r}^{\alpha})^{*}\psi_{\bf r}^{\beta}$ is exponentially small for $\alpha\neq\beta$ (we drop the spin indices for simplicity), and the matrix $R_{\alpha\beta}^{(+)}$ becomes approximately diagonal. Using the same argument, the dominant contribution to $\bar n_{\up}$ also comes from orbital $\alpha$, which gives
\begin{equation}
R^{U}_{\alpha\beta}\approx U\sum_{j}|\psi_{j}^{\alpha}|^{4}\,\delta_{\alpha\beta}=U\,\mathrm{IPR}(\psi^{\alpha})\,\delta_{\alpha\beta},
\label{eq:R_U_IPR_SM}
\end{equation}
and hence
\begin{equation}
\epsilon^{(+)}_{\alpha}\approx\epsilon_{\alpha}+U\,\mathrm{IPR}(\psi^{\alpha}),
\label{eq:charge1_diag_SM}
\end{equation}
which is the diagonal approximation quoted in Eq.~\eqref{eq:charge1_spectrum_simplified} of the main text. To obtain explicit analytical expressions, we must therefore derive the single-particle energies and the inverse participation ratios in the localized regime. To do so, we will employ the locator expansion in the next section.

\subsection{Locator expansion around the atomic limit}

\label{sec:charge1_locator_expansion}

We now perform a locator expansion \cite{PhysRev.109.1492,HAYDOCK198011} around the limit $V=\infty$ (or equivalently $t=0$) for the one-dimensional Aubry-André model.  We will be interested in inspecting the localization properties of the single-particle eigenstates, and in particular in obtaining an analytical expression for their IPR, that we can replace in Eq.$\,$\ref{eq:charge1_diag_SM} to analytically compute the charge +1 excitation energies. We rewrite the spinless Hamiltonian for the one-dimensional Aubry-André  model in the single-particle basis $\ket{j}=c^\dagger_j \ket{0}$, where $c^\dagger_j$ creates an electron at site $j$, as
\begin{equation}
H_{0}=t\sum_{j}(\ket{j}\bra{j+1}+\ket{j+1}\bra{j})+V\sum_{j}v_{j}\ket{j}\bra{j},
\qquad v_{j}=\cos(2\pi\tau j+\phi).
\label{eq:H0_locator_SM}
\end{equation}
For the derivation it is convenient to work with open boundary conditions. In the atomic limit, the highest-energy states of the 
narrow band originate from resonant neighboring sites, so we isolate a two-site cluster $C=\{j,j+1\}$ and partition the chain into sites $L=\{1,\ldots,j-1\}$, $C$, and $R=\{j+2,\ldots,N\}$. In this basis,
\begin{equation}
E-H_{0}=\left(\begin{array}{cccc}
E-H_{LL} & b_{L} & 0 & 0\\
b_{L}^{\dagger} & E-Vv_{j} & -t & 0\\
0 & -t & E-Vv_{j+1} & b_{R}^{\dagger}\\
0 & 0 & b_{R} & E-H_{RR}
\end{array}\right),
\end{equation}
where $b_{L}=-t\ket{j-1}$ and $b_{R}=-t\ket{j+2}$. Equivalently,
\begin{equation}
E-H_{0}=\left(\begin{array}{cc}
E-H_{LR} & u\\
u^{\dagger} & E-H_{C}
\end{array}\right),
\end{equation}
with
\begin{equation}
H_{LR}=\left(\begin{array}{cc}
H_{L} & 0\\
0 & H_{R}
\end{array}\right),
\qquad
H_{C}=\left(\begin{array}{cc}
Vv_{j} & t\\
t & Vv_{j+1}
\end{array}\right),
\qquad
u=\left(\begin{array}{c}
b_{L}\\
b_{R}
\end{array}\right).
\end{equation}
The cluster Green's function is therefore
\begin{equation}
G_{CC}(E)=\bigl(E-H_{C}-\Sigma(E)\bigr)^{-1},
\qquad
\Sigma(E)=u^{\dagger}(E-H_{LR})^{-1}u,
\label{eq:GCC_locator_SM}
\end{equation}
where the self-energy takes the diagonal form
\begin{equation}
\Sigma(E)=\left(\begin{array}{cc}
t^{2}g_{j-1}^{L}(E) & 0\\
0 & t^{2}g_{j+2}^{R}(E)
\end{array}\right).
\label{eq:selfenergy_locator_SM}
\end{equation}
Here
\begin{equation}
g_{j}^{R}(E)=\frac{1}{E-Vv_{j}-t^{2}g_{j+1}^{R}(E)},\qquad g_{N+1}^{R}=0,
\end{equation}
\begin{equation}
g_{j}^{L}(E)=\frac{1}{E-Vv_{j}-t^{2}g_{j-1}^{L}(E)},\qquad g_{0}^{L}=0,
\label{eq:locator_recursions_SM}
\end{equation}
which are the standard left/right locator recursions.

The poles of $G_{CC}(E)$ determine the renormalized energies. Ignoring the self-energy as a first step, one obtains for the poles 
\begin{equation}
E_{\pm}^{0,j}=V\left(\frac{v_{j}+v_{j+1}}{2}\pm\Omega_{j}\right),
\qquad
\Omega_{j}=\sqrt{\left(\frac{\Delta_{1}^{(j)}}{2}\right)^{2}+\left(\frac{t}{V}\right)^{2}},
\label{eq:E0_pm_SM}
\end{equation}
where $\Delta_{i}^{(j)}=v_{j}-v_{j+i}$. Treating the self-energy perturbatively then yields
\begin{equation}
E_{\pm}^{j}=V\epsilon_{\pm}^{0,j}+\frac{t^{2}}{V}\,\delta\epsilon_{\pm}^{j}+\mathcal{O}\!\left(\frac{t^{4}}{V^{3}}\right),
\label{eq:E_pm_locator_SM}
\end{equation}
with $\epsilon_{\pm}^{0,j}=E_{\pm}^{0,j}/V$ and
\begin{equation}
\delta\epsilon_{\pm}^{j}=\frac{1}{2(\epsilon_{\pm}^{0,j}-v_{j-1})}\left(1\pm\frac{\Delta_{1}^{(j)}}{2\Omega_{j}}\right)+\frac{1}{2(\epsilon_{\pm}^{0,j}-v_{j+2})}\left(1\mp\frac{\Delta_{1}^{(j)}}{2\Omega_{j}}\right).
\end{equation}
The lowest narrow band corresponds to the lower branch $E_{-}^{j}$. In the localized regime the ordering of these orbitals is inherited from the atomic limit, and the orbital at filling $\nu=\alpha/N_B$, with $\alpha=0,\cdots,N_B-1$ the narrow-band eigenstate index,  can be associated with a position $j_{\nu}$ satisfying
\begin{equation}
2\pi\tau j_{\nu}=\pi+\frac{\nu\pi}{\ell}+2\pi n,
\qquad \ell=(1-\tau)^{-1},
\label{eq:jnu_SM}
\end{equation}
where $\ell$ is the moir\'e length. Explicitly, this yields
\begin{equation}
\epsilon_{\alpha}\approx E_{-}^{j_{\nu}},\qquad \nu=\alpha/N_{B}.
\end{equation}

To compute the interacting contribution, we also need the eigenfunctions. For a tridiagonal chain, the ratio of Green's functions can be written directly in terms of the locator functions. For $r\ge1$,
\begin{equation}
\frac{G_{j+r,j}(E)}{G_{j,j}(E)}=t^{r}\prod_{s=1}^{r}g_{j+s}^{R}(E),
\qquad
\frac{G_{j-r,j}(E)}{G_{j,j}(E)}=t^{r}\prod_{s=1}^{r}g_{j-s}^{L}(E).
\label{eq:GF_ratio_SM}
\end{equation}
Using the spectral representation
\begin{equation}
G_{ij}(E)=\sum_{\alpha}\frac{\psi_{i}^{\alpha}(\psi_{j}^{\alpha})^{*}}{E-\epsilon_{\alpha}},
\end{equation}
one finds, at a simple pole $E=\epsilon_{\alpha}$,
\begin{equation}
\frac{\psi_{j+r}^{\alpha}}{\psi_{j}^{\alpha}}=t^{r}\prod_{s=1}^{r}g_{j+s}^{R}(\epsilon_{\alpha}),
\qquad
\frac{\psi_{j-r}^{\alpha}}{\psi_{j}^{\alpha}}=t^{r}\prod_{s=1}^{r}g_{j-s}^{L}(\epsilon_{\alpha}).
\label{eq:psi_ratio_SM}
\end{equation}
Evaluating these expressions at $\epsilon_{\alpha}\approx E_{-}^{j_{\nu}}$ generates the full locator expansion for the orbital $\psi^{\alpha}\equiv\psi^{j_{\nu}}$. In practice, the perturbed eigenfunction can be reconstructed recursively from a small set of component ratios. Defining
\begin{equation}
\Delta_{\pm,i}^{(j)}\equiv \epsilon_{\pm}^{0,j}-v_{j+i},
\end{equation}
we first use the cluster eigenvalue equation at $E=E_{\pm}^{j}$,
\begin{equation}
\left(\begin{array}{cc}
E_{\pm}^{j}-Vv_{j}-t^{2}g_{j-1}^{L}(E_{\pm}^{j}) & -t\\
-t & E_{\pm}^{j}-Vv_{j+1}-t^{2}g_{j+2}^{R}(E_{\pm}^{j})
\end{array}\right)
\left(\begin{array}{c}
\psi_{j}^{\alpha}\\
\psi_{j+1}^{\alpha}
\end{array}\right)=0,
\end{equation}
which gives
\begin{equation}
\begin{array}{c}
\dfrac{\psi_{j+1}^{\alpha}}{\psi_{j}^{\alpha}}
=\dfrac{t}{E_{\pm}^{j}-Vv_{j+1}-t^{2}g_{j+2}^{R}(E_{\pm}^{j})}\\
=\dfrac{t}{V\Delta_{\pm,1}^{(j)}}\left[1+\dfrac{t^{2}}{V^{2}\Delta_{\pm,1}^{(j)}}\left(\dfrac{1}{\Delta_{\pm,2}^{(j)}}-\delta\epsilon_{\pm}^{j}\right)\right]+\mathcal{O}\!\left(\dfrac{t^{5}}{V^{5}}\right).
\end{array}
\label{eq:psi_jp1_over_j_SM}
\end{equation}
Similarly, using Eq.~\eqref{eq:psi_ratio_SM} together with the recursion for $g_{j+2}^{R}(E)$, we obtain
\begin{equation}
\begin{array}{c}
\dfrac{\psi_{j+2}^{\alpha}}{\psi_{j+1}^{\alpha}}
=\lim_{E\to E_{\pm}^{j}}t\,g_{j+2}^{R}(E)\\
=\dfrac{t}{V\Delta_{\pm,2}^{(j)}}\left[1+\dfrac{t^{2}}{V^{2}\Delta_{\pm,2}^{(j)}}\left(\dfrac{1}{\Delta_{\pm,3}^{(j)}}-\delta\epsilon_{\pm}^{j}\right)\right]+\mathcal{O}\!\left(\dfrac{t^{5}}{V^{5}}\right).
\end{array}
\label{eq:psi_jp2_over_jp1_SM}
\end{equation}
Repeated use of Eq.~\eqref{eq:psi_ratio_SM} then yields the further right-tail amplitudes
\begin{equation}
\frac{\psi_{j+3}^{\alpha}}{\psi_{j+1}^{\alpha}}=\frac{t^{2}}{V^{2}\Delta_{\pm,2}^{(j)}\Delta_{\pm,3}^{(j)}}+\mathcal{O}\!\left(\frac{t^{4}}{V^{4}}\right),
\qquad
\frac{\psi_{j+4}^{\alpha}}{\psi_{j+1}^{\alpha}}=\frac{t^{3}}{V^{3}\Delta_{\pm,2}^{(j)}\Delta_{\pm,3}^{(j)}\Delta_{\pm,4}^{(j)}}+\mathcal{O}\!\left(\frac{t^{5}}{V^{5}}\right).
\label{eq:psi_right_tail_SM}
\end{equation}
On the left side of the resonant cluster, one analogously finds
\begin{equation}
\begin{array}{c}
\dfrac{\psi_{j-1}^{\alpha}}{\psi_{j}^{\alpha}}=\lim_{E\to E_{\pm}^{j}}t\,g_{j-1}^{L}(E)\\
=\dfrac{t}{V\Delta_{\pm,-1}^{(j)}}\left[1+\dfrac{t^{2}}{V^{2}\Delta_{\pm,-1}^{(j)}}\left(\dfrac{1}{\Delta_{\pm,-2}^{(j)}}-\delta\epsilon_{\pm}^{j}\right)\right]+\mathcal{O}\!\left(\dfrac{t^{5}}{V^{5}}\right),
\end{array}
\label{eq:psi_jm1_over_j_SM}
\end{equation}
with the further left-tail amplitudes
\begin{equation}
\frac{\psi_{j-2}^{\alpha}}{\psi_{j}^{\alpha}}=\frac{t^{2}}{V^{2}\Delta_{\pm,-1}^{(j)}\Delta_{\pm,-2}^{(j)}}+\mathcal{O}\!\left(\frac{t^{4}}{V^{4}}\right),
\qquad
\frac{\psi_{j-3}^{\alpha}}{\psi_{j}^{\alpha}}=\frac{t^{3}}{V^{3}\Delta_{\pm,-1}^{(j)}\Delta_{\pm,-2}^{(j)}\Delta_{\pm,-3}^{(j)}}+\mathcal{O}\!\left(\frac{t^{5}}{V^{5}}\right).
\label{eq:psi_left_tail_SM}
\end{equation}
In practice, we are interested in the lower branch $E_{-}^{j}$, but keeping the $\pm$ notation at this stage makes the derivation transparent. Together with normalization, Eqs.~\eqref{eq:psi_jp1_over_j_SM}--\eqref{eq:psi_left_tail_SM} determine the full perturbed eigenfunction that we used for the analytical calculations shown in the main text, in Fig.$\,$\ref{fig:2}.

The resulting closed-form expression for the IPR, $\mathrm{IPR}(\psi^{\alpha})=\sum_{j}|\psi_{j}^{\alpha}|^{4}$, is lengthy, so below we will quote the asymptotic forms relevant for the lowest and highest-energy narrow-band states, respectively $\nu=0$ and $\nu=1$.

Introducing $\lambda=t/V$, one finds near $\nu=0$,
\begin{equation}
\begin{array}{c}
\mathrm{IPR}(\psi^{\nu\approx0})\approx1-\frac{\lambda^{2}\csc^{4}(\pi/\ell)\left(\ell^{2}-2\pi^{2}\nu^{2}+3\pi^{2}\nu^{2}\csc^{2}(\pi/\ell)\right)}{\ell^{2}}\\
\approx1-\frac{\lambda^{2}\ell^{4}}{\pi^{4}}(1+3\nu^{2}),\qquad (\lambda\ll1,\;\nu\approx0,\;\ell\gg1),
\end{array}
\label{eq:IPR_nu0_SM}
\end{equation}
so the lowest-energy eigenstates depart smoothly from atomic states (with $\mathrm{IPR}\approx1$).

\begin{figure}[t]
    \centering
    \includegraphics[width=\columnwidth]{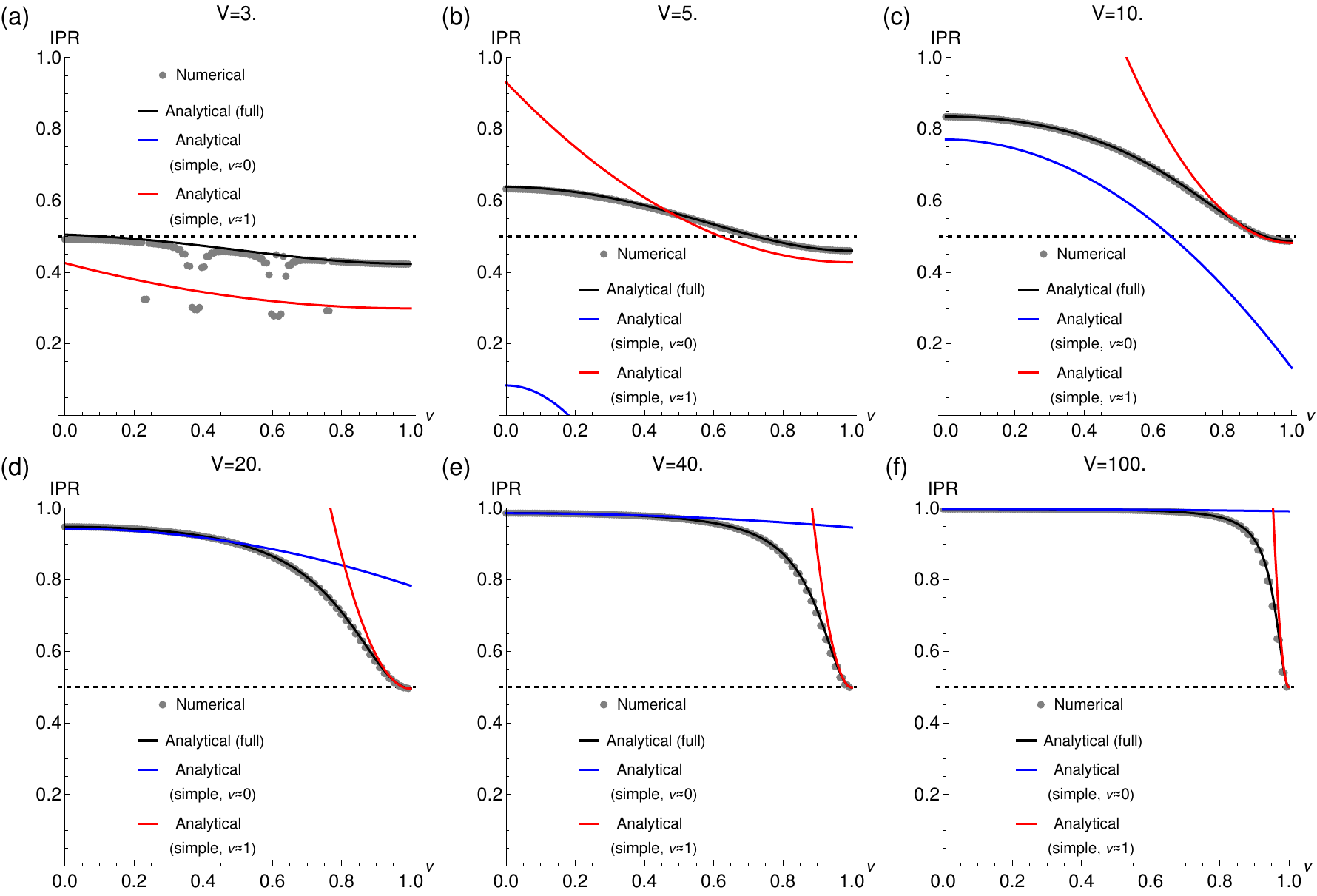}
    \caption{\textrm{IPR} of the single-particle eigenstate as a function of filling $\nu$  of the lowest narrow band. Different panels correspond to different V indicated in each panel. The numerical results were obtained for $L=953$, with system size N=953, an approximant of $\tau=\frac{47+\sqrt{5}}{58}$. The 'Analytical (full)' curves were obtained using the full locator expansion. The remaining (simple) analytical curves correspond to the simplified expressions in Eqs.$\,$\ref{eq:IPR_nu0_SM} and \ref{eq:IPR_nu1_SM}. }
    \label{fig:IPR_vs_nu_SM}
\end{figure}

For the highest-energy states of the narrow band, the exact local resonance $\Delta_{1}^{(j_{1})}=0$ makes the expansion non-analytic at $\lambda=0$. The origin of this non-analyticity is the factor
\begin{equation}
\Omega_{j_{\nu}}=\sqrt{\left(\frac{\Delta_{1}^{(j_{\nu})}}{2}\right)^{2}+\lambda^{2}},
\end{equation}
which enters both the energies and the eigenvectors. Near $\nu=1$, the relevant small parameter is therefore not $\lambda$ alone, but rather the ratio $(\nu-1)/\lambda$. In the regime $(\nu-1)/\lambda\ll1$, one obtains
\begin{equation}
\begin{array}{c}
\mathrm{IPR}(\psi^{\nu\approx1})\approx\frac{1}{2}+\frac{\pi^{2}(\nu-1)^{2}}{2\ell^{2}}\left[\frac{1}{\lambda^{2}}\sin^{2}\!\left(\frac{\pi}{\ell}\right)-\frac{1}{8}\csc^{2}\!\left(\frac{\pi}{\ell}\right)\sec^{2}\!\left(\frac{\pi}{\ell}\right)\right]\\
-\frac{\lambda^{2}}{16}\csc^{4}\!\left(\frac{\pi}{\ell}\right)\sec^{2}\!\left(\frac{\pi}{\ell}\right),\qquad \big[(\nu-1)/\lambda\ll1,\;\nu\approx1\big],
\end{array}
\label{eq:IPR_nu1_SM}
\end{equation}
The leading correction is the $\mathcal{O}(1/\lambda^{2})$ term proportional to $(\nu-1)^{2}\sin^{2}(\pi/\ell)/\ell^{2}$, which quantifies how rapidly the IPR departs from the perfectly resonant value $1/2$ as one moves away from the band edge. The locator expansion therefore explains the monotonic decrease of the IPR across the narrow band: for large $V$, the eigenstates at the bottom of the band are almost atomic, while the top of the band is formed by nearly perfect two-site resonances.
In Fig.$\,$\ref{fig:IPR_vs_nu_SM}, we show a detailed comparison between the analytical results here derived and the exact numerical results, for different $V$.

\subsection{Analytical results for charge +1 excitations}

\label{sec:charge1_analytical}

Substituting the full locator-expanded eigenfunctions into Eq.~\eqref{eq:R_U_IPR_SM} yields a long expression which we omit here. We instead provide simplified expressions below for $\nu=0$ and $\nu=1$,
\begin{equation}
E^{U}_{\nu=0}\approx U\left(1+\frac{\csc^{4}(\pi\tau)}{V^{2}}\right)^{-1},
\qquad
E^{U}_{\nu=1}\approx U\left(2+\frac{4}{V^{2}[\cos(\pi\tau)-\cos(3\pi\tau)]^{2}}\right)^{-1}.
\label{eq:EU_edges_SM}
\end{equation}
At large $V$, these approach $U$ and $U/2$, respectively, as predicted by Eqs.$\,$\ref{eq:IPR_nu0_SM} and \ref{eq:IPR_nu1_SM}.

The full expressions for the renormalized single-particle energies are also long, and therefore we present below the simplified results,
\begin{equation}
\begin{array}{c}
\epsilon_{\nu=0}\approx -V\cos(\pi\tau)\left[\cos(\pi\tau)+\sin\!\left(\frac{\pi\tau}{2}\right)\right]+\sqrt{1+V^{2}\cos^{2}\!\left(\frac{\pi\tau}{2}\right)\sin^{2}(\pi\tau)}-\sqrt{1+V^{2}\sin^{4}(\pi\tau)},\\
\epsilon_{\nu=1}\approx -1+V\cos(\pi\tau)\left[1-\sin\!\left(\frac{\pi\tau}{2}\right)\right]+\sqrt{1+V^{2}\cos^{2}\!\left(\frac{\pi\tau}{2}\right)\sin^{2}(\pi\tau)}.
\end{array}
\label{eq:epsilon_edges_SM}
\end{equation}
In the asymptotic limit $V\to\infty$ and $\tau\to1$, these reduce to $\epsilon_{\nu=0}\to-(\pi^{2}/8)V(\tau-1)^{2}$ and $\epsilon_{\nu=1}\to(3\pi^{2}/8)V(\tau-1)^{2}$.

\begin{figure}[t]
    \centering
    \includegraphics[width=\columnwidth]{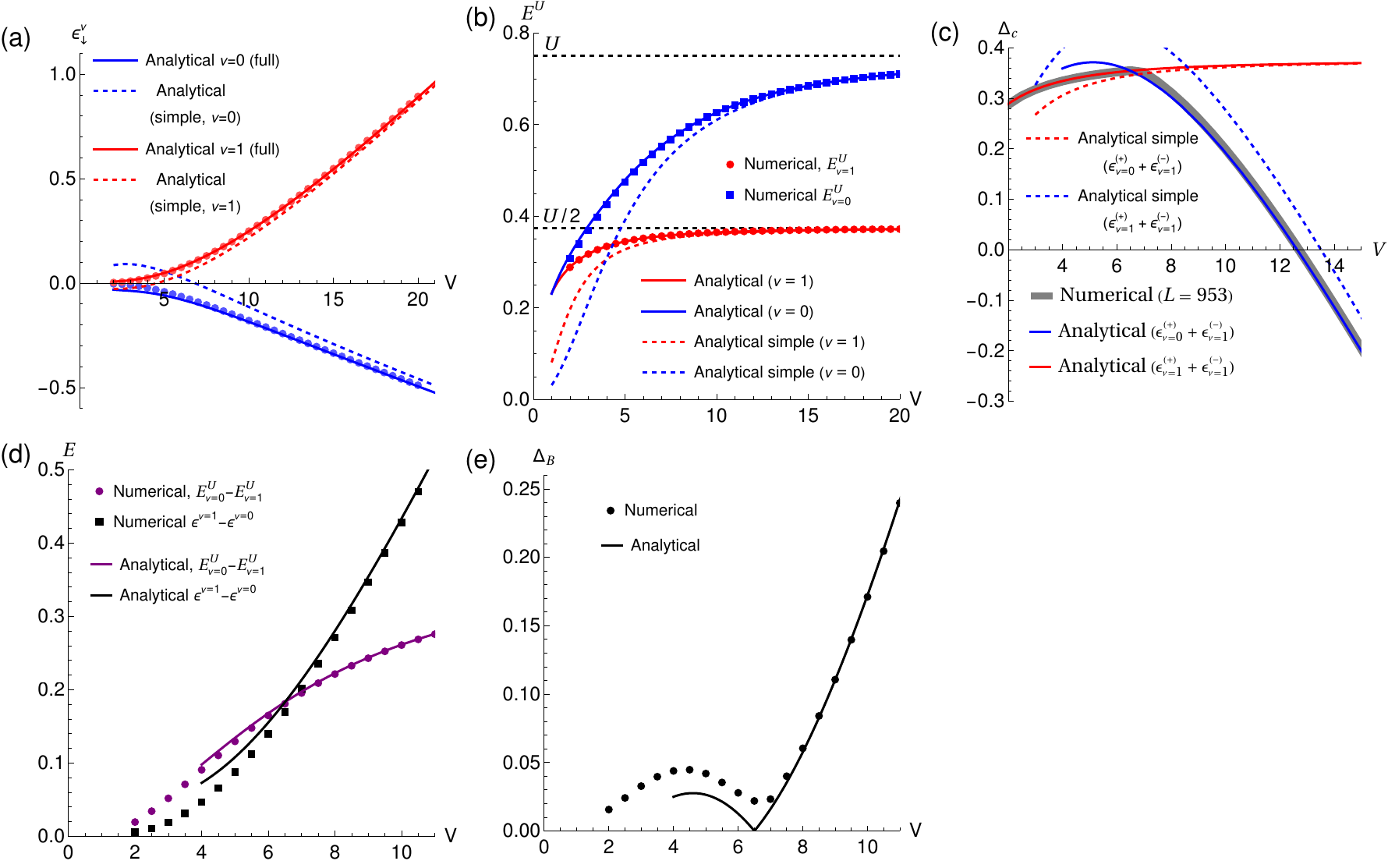}
    \caption{
Charge +1 excitation results for $U=0.75$ and $\tau=\frac{47+\sqrt{5}}{58}$. The numerical results were obtained for $N=1542$. (a) Single-particle energies of the eigenstates at filling $\nu=0$ and $\nu=1$ of the narrow-band. The numerically exact results are compared with the full analytical solution from Eq.$\,$\ref{eq:E_pm_locator_SM} and the simplified expressions in Eq.$\,$\ref{eq:epsilon_edges_SM}. (b) Lowest and highest eigenvalues of $R_{\alpha\beta}^{U}$, together with the analytical results from Eq.$\,$\ref{eq:charge1_diag_SM}. The full analytical curves use the full locator expansion derived in Sec.$\,$\ref{sec:charge1_locator_expansion}, while the simplified curves correspond to Eq.$\,$\ref{eq:EU_edges_SM}. (c) Exact charge gap $\Delta_c \equiv \textrm{min}_{\{\alpha,\beta \}}( \epsilon^{(+)}_{\alpha} + \epsilon^{(-)}_{\beta})$, together with the analytical predictions obtained from the charge +1 excitations at $\nu=0$ and $\nu=1$ and the charge -1 excitation at $\nu=1$. (d) Single-particle bandwidth and bandwidth of the spectrum of the $R^{U}$ matrix. (e) Full bandwidth of the charge excitations. The analytical curve is obtained from the absolute difference between the charge-excitation energies at $\nu=0$ and $\nu=1$.}
    \label{fig:charge1_SM}
\end{figure}

In Fig.$\,$\ref{fig:charge1_SM} we compare the exact numerical charge-excitation results with both the simplified analytical expressions shown above and the analytical results using the full locator-expansion calculation in Sec.$\,$\ref{sec:charge1_locator_expansion}. Panels (a) and (b) show that the full analytical results  reproduce the single-particle and interaction charge +1 energies very accurately over a broad range of $V$, while the simplified expressions capture the correct large-$V$ asymptotics and capture well the qualitative behaviour at smaller $V$. Panels (c)--(e) reveal an intermediate-$V$ crossover: the lowest charge excitation is controlled by the $\nu=1$ charge +1 excitation at smaller $V$, but by the $\nu=0$ charge +1 state deeper in the localized regime. This crossover is accompanied by a pronounced minimum in the bandwidth of the charge +1 spectrum. Because the off-diagonal matrix elements of $R^{U}_{\alpha\beta}$, although small, are not exactly zero, the bandwidth has a minimum but remains finite around the crossover, as shown in panel (e).

\pagebreak

\section{Spin-flip spectrum and effective magnon Hamiltonian}

In the main text we showed that one-spin-flip excitations are described by a projected two-particle problem and that, in the localized regime, a narrow collective magnon branch sets the lowest-energy excitations. In this section we provide a compact derivation of the exact projected two-particle Hamiltonian, show in detail why the number of magnon bound modes equals the number of orbitals in the narrow band in the large-$V$ limit, and then derive the effective low-energy Hamiltonian for the magnons.

\subsection{Effective two-particle spin-flip Hamiltonian}

\label{sec:spinflip_Heff}

One-spin-flip excitations above the fully polarized state are spanned by
\begin{equation}
\ket{\alpha,\beta}=S_{\alpha\beta}\ket{\Uparrow},
\qquad
S_{\alpha\beta}\equiv\gamma_{\down,\alpha}^{\dagger}\gamma_{\up,\beta},
\end{equation}
where $\alpha$ labels the added spin-$\down$ particle and $\beta$ the removed spin-$\up$ particle. The exact projected spin-flip problem therefore reduces to the two-particle matrix
\begin{equation}
\mathcal{H}^{\textrm{\ensuremath{\down\up}}}_{(\lambda\delta,\alpha\beta)}
\equiv
\bra{\Uparrow}S_{\lambda\delta}^{\dagger}\bar H S_{\alpha\beta}\ket{\Uparrow}
-E_{\Uparrow}\delta_{\lambda\alpha}\delta_{\delta\beta}.
\label{eq:magnon_H_SM_def}
\end{equation}

The non-interacting contribution is immediate from the commutators
\begin{equation}
[\bar H_{0},\gamma_{\down,\alpha}^{\dagger}]=\epsilon_{\alpha}\gamma_{\down,\alpha}^{\dagger},
\qquad
[\bar H_{0},\gamma_{\up,\beta}]=-\epsilon_{\beta}\gamma_{\up,\beta},
\end{equation}
together with $\bar H_{0}\ket{\Uparrow}=E_{\Uparrow}\ket{\Uparrow}$. Therefore
\begin{equation}
\bar H_{0}\gamma_{\down,\alpha}^{\dagger}\gamma_{\up,\beta}\ket{\Uparrow}
=(E_{\Uparrow}+\epsilon_{\alpha}-\epsilon_{\beta})\gamma_{\down,\alpha}^{\dagger}\gamma_{\up,\beta}\ket{\Uparrow},
\end{equation}
which yields
\begin{equation}
\bra{\Uparrow}S_{\lambda\delta}^{\dagger}\bar H_{0}S_{\alpha\beta}\ket{\Uparrow}
-E_{\Uparrow}\delta_{\lambda\alpha}\delta_{\delta\beta}
=(\epsilon_{\alpha}-\epsilon_{\beta})\delta_{\lambda\alpha}\delta_{\delta\beta}.
\label{eq:magnon_H0_SM}
\end{equation}

For the interaction term, since $\bar H_{U}\ket{\Uparrow}=0$, we can write
\begin{equation}
\bra{\Uparrow}S_{\lambda\delta}^{\dagger}\bar H_{U}S_{\alpha\beta}\ket{\Uparrow}
=\bra{\Uparrow}\gamma_{\up,\delta}^{\dagger}\gamma_{\down,\lambda}
[\bar H_{U},\gamma_{\down,\alpha}^{\dagger}\gamma_{\up,\beta}]\ket{\Uparrow}.
\end{equation}
Expanding the commutator gives two contributions. The first one is the same matrix element already encountered in the charge +1 problem:
\begin{equation}
\bra{\Uparrow}\gamma_{\up,\delta}^{\dagger}\gamma_{\down,\lambda}
[\bar H_{U},\gamma_{\down,\alpha}^{\dagger}]\gamma_{\up,\beta}\ket{\Uparrow}
=R^{U}_{\lambda\alpha}\delta_{\delta\beta}.
\end{equation}
The second one comes from the interaction acting on the removed spin-$\up$ electron and produces the direct particle-hole binding term,
\begin{equation}
\bra{\Uparrow}\gamma_{\up,\delta}^{\dagger}\gamma_{\down,\lambda}
\gamma_{\down,\alpha}^{\dagger}[\bar H_{U},\gamma_{\up,\beta}]\ket{\Uparrow}
=-U\,V_{\up\down,\beta\lambda\alpha\delta}.
\end{equation}
Combining both pieces with Eq.~\eqref{eq:magnon_H0_SM}, we obtain the exact projected two-particle Hamiltonian
\begin{equation}
\mathcal{H}^{\textrm{\ensuremath{\down\up}}}_{(\lambda\delta,\alpha\beta)}
=\left[(\epsilon_{\alpha}-\epsilon_{\beta})\delta_{\lambda\alpha}+R^{U}_{\lambda\alpha}\right]\delta_{\delta\beta}
-U\,V_{\up\down,\beta\lambda\alpha\delta}.
\label{eq:magnon_H_SM_exact}
\end{equation}
This completes the derivation of Eq.~\eqref{eq:magnon_H_full} in the main text. The two terms also have a transparent meaning. The first term provides the contribution from independent (unbound) particle-hole excitations, while the second term defines the particle-hole interaction/binding matrix responsible for the creation of magnon bound modes.

\begin{figure}[h!]
    \centering
    \includegraphics[width=\columnwidth]{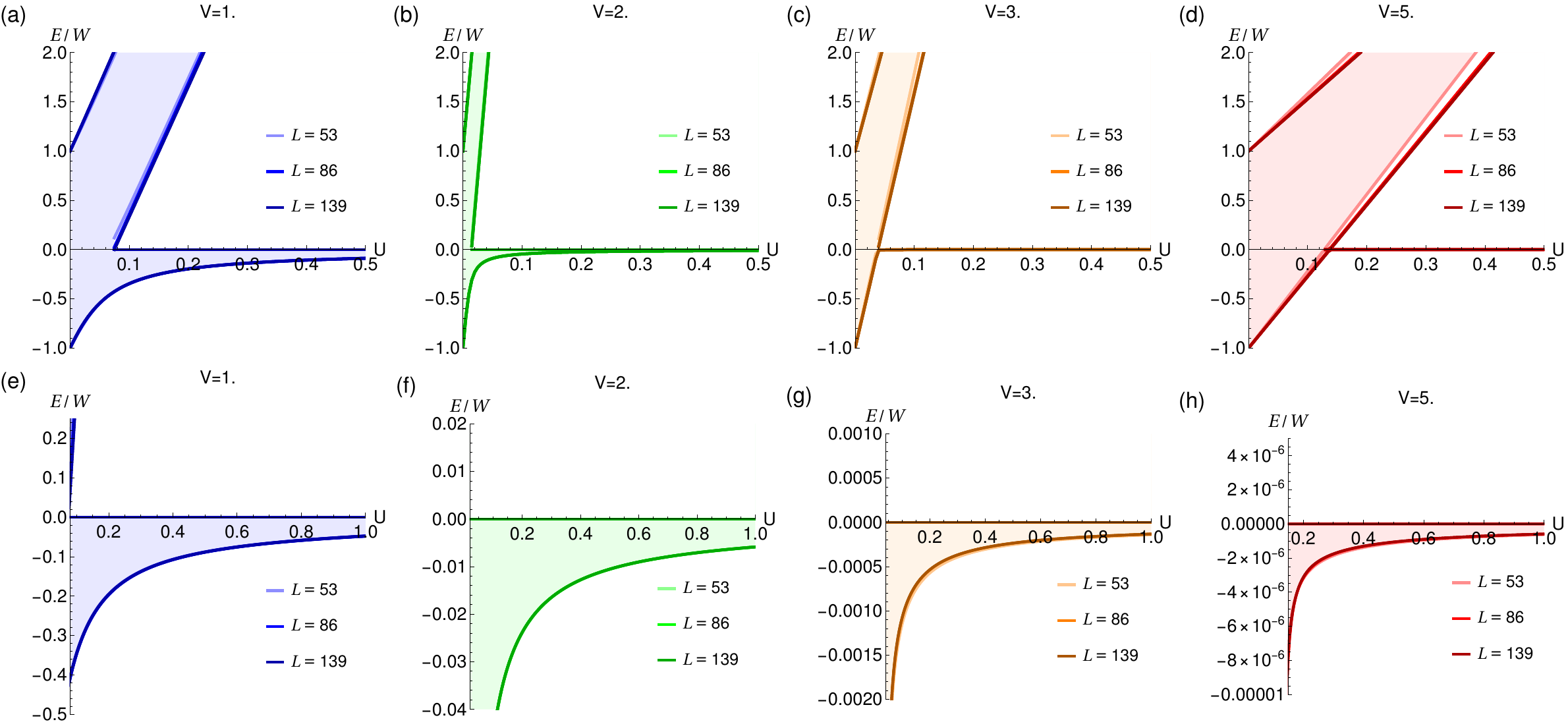}
    \caption{Spin-flip spectrum obtained by diagonalizing Eq.$\,$\ref{eq:magnon_H_SM_exact} as function of $U$, for different system sizes. The results are for the 1D Aubry-André model with $t_2=t_3=0$ and different values of $V$ indicated in each panel. (a-d) Results for the magnon modes and continuum spectrum. (e-h) Close-up on the magnon spectrum.}
    \label{fig:magnon_spec_finite_size_SM}
\end{figure}

\begin{figure}[h!]
    \centering
    \includegraphics[width=\columnwidth]{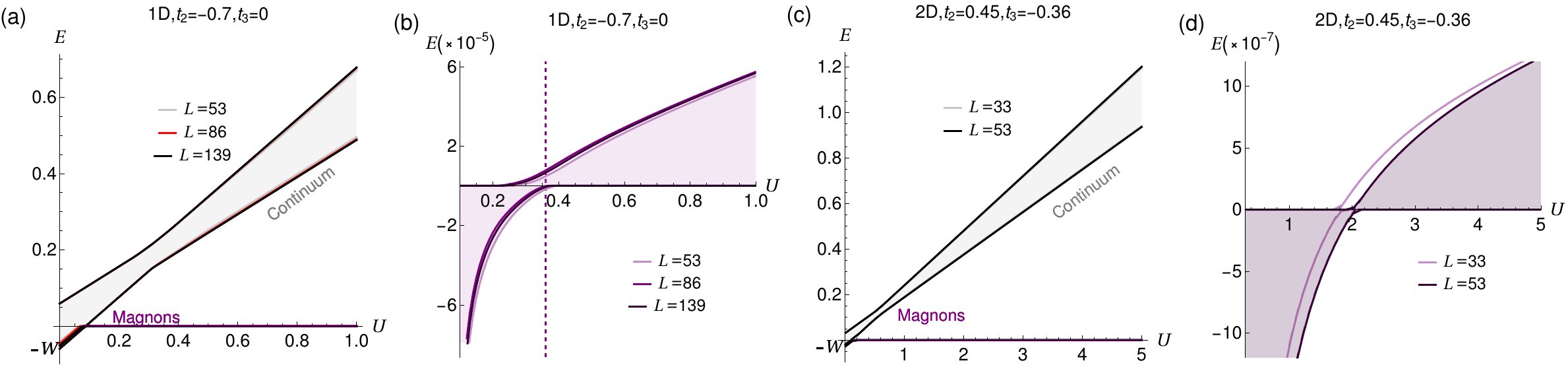}
    \caption{Example spin-flip spectra obtained by diagonalizing Eq.$\,$\ref{eq:magnon_H_SM_exact} as function of $U$ in 1D (a,b) and 2D (c,d) for different system sizes, for the parameters indicated on top of each panel. (a,c) show the full spin-flip spectrum, while (b,d) show a close-up at the magnon spectrum.}
    \label{fig:magnon_spec_finite_1D_and_2D_SM}
\end{figure}

In Figs.$\,$\ref{fig:magnon_spec_finite_size_SM} and \ref{fig:magnon_spec_finite_1D_and_2D_SM} we show some examples demonstrating the robustness of the magnon spectrum obtained by diagonalizing the Hamiltonian in Eq.$\,$\ref{eq:magnon_H_SM_exact} for different sizes.

\subsection{Bound mode counting}

\label{sec:spinflip_bound_counting}

A useful way of organizing Eq.~\eqref{eq:magnon_H_SM_exact} is to separate the continuum matrix from the binding term. Writing
\begin{equation}
\mathcal{H}^{\down\up}_{(\lambda\delta,\alpha\beta)}=R_{(\lambda,\delta),(\alpha,\beta)}-U\sum_{j}\mathcal{V}_{(\lambda,\delta),j}\mathcal{V}^{\dagger}_{j,(\alpha,\beta)},
\end{equation}
with
\begin{equation}
\mathcal{V}_{(\lambda,\delta),j}=\psi_{j}^{\delta}\left(\psi_{j}^{\lambda}\right)^{*},
\qquad
R_{(\lambda,\delta),(\alpha,\beta)}= \left[(\epsilon_{\alpha}-\epsilon_{\beta})\delta_{\lambda\alpha}+R^{U}_{\lambda\alpha}\right]\delta_{\delta\beta}
,
\end{equation}
we can express the full spin-flip Hamiltonian as
\begin{equation}
\mathcal{H}^{\down\up}=R-U\mathcal{V}\mathcal{V}^{\dagger}.
\label{eq:magnon_R_minus_VV_SM}
\end{equation}
States inside the particle-hole continuum satisfy $\det(R-E)=0$, whereas the collective modes that are well isolated from the continuum are obtained from
\begin{equation}
\det\!\left(1-\mathcal{V}^{\dagger}\frac{U}{R-E}\mathcal{V}\right)=0.
\label{eq:magnon_bound_condition_SM}
\end{equation}
This reduces the bound-state problem from an $N_{B}^{2}\times N_{B}^{2}$ matrix to an $N\times N$ one, with $N=L^d$, implying that there are at most $N$ bound eigenmodes. However, as discussed in the main text, we only see a single low energy magnon branch with $N_B$ eigenmodes.
This counting becomes particularly transparent for large $V$, with $U\gg W$. In this limit, the denominator in Eq.~\eqref{eq:magnon_bound_condition_SM} is dominated by the interaction term and the eigenstates are very localized, so that effectively $R_{\lambda\alpha}\approx U\delta_{\lambda\alpha}$. Equation~\eqref{eq:magnon_bound_condition_SM} then gives
\begin{equation}
\mathcal{V}^{\dagger}\frac{U}{R-E}\mathcal{V}\approx \frac{U}{U-E}|\Pi|^{2},
\qquad
\Pi_{ij}=\sum_{\alpha}\psi_{i}^{\alpha}(\psi_{j}^{\alpha})^{*}.
\label{eq:magnon_atomic_bound_SM}
\end{equation}
For large $V$, $\Pi$ is approximately diagonal and only has support on the $N_{B}$ sites where the narrow-band eigenstates are localized (within each moiré cell). Therefore $|\Pi|^{2}$ has rank $N_{B}$ up to exponentially suppressed corrections in $\ell/\xi$, where $\ell$ is the moiré length and $\xi$ is the localization length. Therefore there are exactly $N_{B}$ collective bound modes. At strictly infinite $V$ these modes have exactly zero energy and reduce to the local spin flips $\gamma_{\down,\alpha}^{\dagger}\gamma_{\up,\alpha}\ket{\Uparrow}$. At finite but large $V$, they hybridize and acquire a finite bandwidth.

\subsection{Effective magnon Hamiltonian}

\label{sec:effective_magnon_H}

We now derive the effective Hamiltonian acting within the low-energy magnon manifold. As discussed in the main text, in the localized regime and for $U>U^{*}$ the relevant low-energy subspace is spanned by the local spin-flip states $\ket{\alpha,\alpha}$. We therefore introduce the projectors
\begin{equation}
P=\sum_{\alpha}\ket{\alpha,\alpha}\bra{\alpha,\alpha},
\qquad
Q=1-P,
\end{equation}
where $Q$ projects onto the non-local particle-hole configurations $\ket{\lambda,\delta}$ with $\lambda\neq\delta$. Integrating out this high-energy sector gives the exact Feshbach-Schur complement
\begin{equation}
H_{\mathrm{eff}}(E)=H_{PP}-H_{PQ}(H_{QQ}-E)^{-1}H_{QP}.
\label{eq:magnon_Heff_SM}
\end{equation}
We are interested in the magnon branch that evolves continuously from the zero-energy manifold at $V=\infty$, so throughout this subsection we set $E=0$ in Eq.~\eqref{eq:magnon_Heff_SM}.

We will follow the notation of the main text, using the definitions
\begin{equation}
[\rho_{\alpha\beta}]_{\mathbf r}=(\psi^{\alpha}_{\mathbf r})^{*}\psi^{\beta}_{\mathbf r},
\qquad
\Delta_{\alpha\beta}=\epsilon_{\alpha}-\epsilon_{\beta}.
\label{eq:magnon_notation_SM}
\end{equation}
Additionally, as in the main text, we assume for simplicity that the Hamiltonian is $SU(2)$-symmetric and therefore suppress the spin labels in the single-particle eigenfunctions. In what follows, we will also use that the Hamiltonians studied in this work are real so the eigenfunctions can be chosen real (but of course the derivation can be straightforwardly generalized). 
In the localized phase, every inter-orbital product between different moir\'e cells brings a factor of order $e^{-\ell/\xi}$, where once again $\ell$ is the moiré length and $\xi$ is the localization length. Because of this, we can greatly simplify the effective Hamiltonian by keeping only the terms to lower-order in powers of $e^{-\ell/\xi}$, as we will show below.

Before evaluating individual matrix elements, let us note that $SU(2)$ symmetry fixes the diagonal entries of $H_{\mathrm{eff}}$ once the off-diagonal ones are known. The uniform local spin flip mode
\begin{equation}
\ket{\psi_{0}}=\sum_{\alpha}\ket{\alpha,\alpha}=\left(\sum_{\alpha}\gamma_{\down,\alpha}^{\dagger}\gamma_{\up,\alpha}\right)\ket{\Uparrow}
\end{equation}
is obtained by acting with the total spin-lowering operator on $\ket{\Uparrow}$, and therefore must remain a zero mode of the effective Hamiltonian. One can check directly from Eq.~\eqref{eq:magnon_H_SM_exact} that
\begin{equation}
\sum_{\beta}(H_{PP})_{\alpha\beta}=0,
\end{equation}
and similarly
\begin{equation}
\sum_{\beta}(H_{QP})_{(\lambda\delta),(\beta\beta)}
=R^{U}_{\lambda\delta}-U\sum_{\mathbf r}\sum_{\beta}|\psi^{\beta}_{\mathbf r}|^{2}(\psi^{\lambda}_{\mathbf r})^{*}\psi^{\delta}_{\mathbf r}=0,
\end{equation}
which implies $\sum_{\beta}\Sigma_{\alpha\beta}=0$ and therefore
\begin{equation}
\sum_{\beta}(H_{\mathrm{eff}})_{\alpha\beta}=0.
\end{equation}
Hence
\begin{equation}
(H_{\mathrm{eff}})_{\alpha\alpha}=-\sum_{\beta\neq\alpha}(H_{\mathrm{eff}})_{\alpha\beta},
\end{equation}
so it is sufficient to compute the off-diagonal matrix elements explicitly and the diagonal ones then follow automatically.

We start from the direct projection of Eq.~\eqref{eq:magnon_H_SM_exact} into the $P$ subspace. For matrix elements between two local spin flips we obtain
\begin{equation}
\begin{aligned}
(H_{PP})_{\alpha\beta}
&=\bra{\alpha,\alpha}\mathcal H^{\down\up}\ket{\beta,\beta}\\
&=R^{U}_{\alpha\alpha}\delta_{\alpha\beta}-U\,V_{\up\down,\beta\alpha\beta\alpha}\\
&=U\sum_{\lambda\neq\alpha}\|\rho_{\alpha\lambda}\|^{2}\,\delta_{\alpha\beta}-U\,\|\rho_{\alpha\beta}\|^{2}(1-\delta_{\alpha\beta}).
\end{aligned}
\label{eq:magnon_HPP_SM}
\end{equation}
In particular, for $\alpha\neq\beta$ the direct coupling is simply
\begin{equation}
(H_{PP})_{\alpha\beta}=-U\|\rho_{\alpha\beta}\|^{2}  \,\,\, (\alpha\neq \beta).
\label{eq:magnon_HPP_offdiag_SM}
\end{equation}
This is the first, negative semi-definite term appearing in Eq.~\eqref{eq:magnon_main_text} in the main text.

We next evaluate the virtual correction
\begin{equation}
\Sigma\equiv H_{PQ}H_{QQ}^{-1}H_{QP},
\qquad \textrm{with }\,
(H_{\mathrm{eff}})_{\alpha\beta}=(H_{PP})_{\alpha\beta}-\Sigma_{\alpha\beta}.
\label{eq:magnon_Sigma_def_SM}
\end{equation}
Given a state $\ket{\alpha,\alpha}$, $H_{PQ}$ will couple it to states $\ket{\alpha,\beta}$ and $\ket{\beta,\alpha}$ with $\beta\neq\alpha$. In general, it can also couple to states  $\ket{\gamma,\beta}$, with $\gamma \neq \beta \neq \alpha$, but these will involve overlaps of eigenstates living in three different moiré cells (instead of two), which will be higher-order in powers of $e^{-\ell/\xi}$. Additionally, we can restrict $\alpha$ and $\beta$ to orbitals at \textit{nearest-neighbor moiré cells}, since more distant orbitals involve higher powers of $e^{-\ell/\xi}$. In what follows, we will employ these approximations and keep track of the error in powers of $e^{-\ell/\xi}$.  

Using Eq.~\eqref{eq:magnon_H_SM_exact}, we have
\begin{equation}
\begin{aligned}
\bra{\alpha,\alpha}H_{PQ}\ket{\alpha,\beta}
&=\bra{\alpha,\alpha}\mathcal H^{\down\up}\ket{\alpha,\beta}\\
&=\left[(\epsilon_{\alpha}-\epsilon_{\beta})\delta_{\alpha\alpha}+R^{U}_{\alpha\alpha}\right]\delta_{\alpha\beta}
-U\,V_{\up\down,\beta\alpha\alpha\alpha}\\
&=-U\sum_{\mathbf r}(\psi^{\beta}_{\mathbf r})^{*}\psi^{\alpha}_{\mathbf r}|\psi^{\alpha}_{\mathbf r}|^{2} \,\, (\alpha\neq\beta)\\
&=-U\,\rho_{\alpha\alpha}\cdot\rho_{\alpha\beta}  \,\, (\alpha\neq\beta).
\end{aligned}
\label{eq:HPQ_alphabeta_SM}
\end{equation}
This relation is exact for $\alpha\neq\beta$: there is no additional localized-expansion approximation for this term.
This is no longer the case for the second coupling, for which we find
\begin{equation}
\bra{\alpha,\alpha}H_{PQ}\ket{\beta,\alpha}=U\sum_{\lambda\neq\alpha}V_{\up\down,\lambda\alpha\beta\lambda}
=U\,\rho_{\beta\beta}\cdot\rho_{\alpha\beta}+\mathcal{O}(e^{-2\ell/\xi}),
\label{eq:HPQ_betaalpha_SM}
\end{equation}
Here, the leading term comes from $\lambda=\beta$, since every term with $\lambda\neq\beta$ contains at least one additional inter-moiré overlap and is therefore suppressed by an extra factor $e^{-\ell/\xi}$. This asymmetry between $\ket{\alpha,\beta}$ and $\ket{\beta,\alpha}$ can be understood by recalling that the first index in the basis represents an electron with spin-down and the second one represents a hole with spin-up. Hermiticity then gives the leading-order $H_{QP}$ matrix elements,
\begin{equation}
\bra{\alpha,\beta}H_{QP}\ket{\beta,\beta}=U\,\rho_{\alpha\alpha}\cdot\rho_{\alpha\beta}+\mathcal{O}(e^{-2\ell/\xi}),
\end{equation}
\begin{equation}
\bra{\beta,\alpha}H_{QP}\ket{\beta,\beta}=-U\,\rho_{\beta\beta}\cdot\rho_{\alpha\beta}.
\end{equation}
The first of these inherits the same localized-expansion approximation as Eq.~\eqref{eq:HPQ_betaalpha_SM} with $\alpha\leftrightarrow\beta$, whereas the second one is exact for the same reason as Eq.~\eqref{eq:HPQ_alphabeta_SM}. 

For the same two intermediate states, the relevant diagonal $Q$-space matrix elements follow from the same exact Hamiltonian:
\begin{equation}
\begin{aligned}
\bra{\alpha,\beta}H_{QQ}\ket{\alpha,\beta}
&=\Delta_{\alpha\beta}+R^{U}_{\alpha\alpha}-U\,V_{\up\down,\beta\alpha\alpha\beta}\\
&=\Delta_{\alpha\beta}+U\sum_{\lambda\neq\beta}\|\rho_{\alpha\lambda}\|^{2}\\
&=\Delta_{\alpha\beta}+U\|\rho_{\alpha\alpha}\|^{2}+\mathcal{O}(e^{-\ell/\xi}),
\end{aligned}
\end{equation}
\begin{equation}
\begin{aligned}
\bra{\beta,\alpha}H_{QQ}\ket{\beta,\alpha}
&=\Delta_{\beta\alpha}+R^{U}_{\beta\beta}-U\,V_{\up\down,\alpha\beta\beta\alpha}\\
&=\Delta_{\beta\alpha}+U\sum_{\lambda\neq\alpha}\|\rho_{\beta\lambda}\|^{2}\\
&=\Delta_{\beta\alpha}+U\|\rho_{\beta\beta}\|^{2}+\mathcal{O}(e^{-\ell/\xi}).
\end{aligned}
\end{equation}
In each case the explicit interaction term cancels the corresponding $\lambda=\beta$ or $\lambda=\alpha$ contribution inside $R^{U}$, leaving a sum whose dominant piece is the local $\lambda=\alpha$ or $\lambda=\beta$ term. The off-diagonal element inside this reduced two-state $Q$ block is
\begin{equation}
\bra{\alpha,\beta}H_{QQ}\ket{\beta,\alpha}=-U\,V_{\up\down,\alpha\alpha\beta\beta}=\mathcal{O}(e^{-\ell/\xi}),
\end{equation}
and couplings from these states to any other $Q$ configuration contain at least one additional inter-orbital overlap. Consequently,  replacing $H_{QQ}$ by the diagonal contributions above includes all contributions up to order $\mathcal{O}(e^{-\ell/\xi})$.

Substituting all the leading matrix elements into Eq.~\eqref{eq:magnon_Sigma_def_SM}, we obtain for $\alpha\neq\beta$
\begin{equation}
\Sigma_{\alpha\beta}
=
-U\left[
\frac{(\rho_{\alpha\alpha}\cdot\rho_{\alpha\beta})^{2}}{\Delta_{\alpha\beta}/U+\|\rho_{\alpha\alpha}\|^{2}}
+
\frac{(\rho_{\beta\beta}\cdot\rho_{\alpha\beta})^{2}}{\Delta_{\beta\alpha}/U+\|\rho_{\beta\beta}\|^{2}}
\right]
+\mathcal{O}(e^{-2\ell/\xi}).
\label{eq:magnon_Sigma_SM}
\end{equation}
Combining Eqs.~\eqref{eq:magnon_HPP_offdiag_SM} and \eqref{eq:magnon_Sigma_SM}, the off-diagonal matrix elements of the effective Hamiltonian become
\begin{equation}
(H_{\mathrm{eff}})_{\alpha\beta}=\frac{J_{\alpha\beta}}{2},
\qquad \alpha\neq\beta,
\label{eq:J_ab_Heff}
\end{equation}
where we defined the effective exchange couplings $J_{\alpha\beta}$ as
\begin{equation}
J_{\alpha\beta}
\approx
-2U\|\rho_{\alpha\beta}\|^{2}\Bigg[
1
-\frac{\cos^{2}\theta_{\alpha\beta}}{1+\frac{\Delta_{\alpha\beta}}{U\|\rho_{\alpha\alpha}\|^{2}}}
-\frac{\cos^{2}\theta_{\beta\alpha}}{1+\frac{\Delta_{\beta\alpha}}{U\|\rho_{\beta\beta}\|^{2}}}
\Bigg]
+\mathcal{O}(e^{-2\ell/\xi}).
\label{eq:magnon_exchange_SM}
\end{equation}
The first term inside the brackets comes from the direct projection into the local-spin-flip manifold, while the second and third terms come from virtual excursions into the non-local particle-hole sector. To see explicitly why these off-diagonal matrix elements are naturally interpreted as exchange couplings, we define the spin operators

\begin{equation}
  \mathbf S_{\alpha}
=
\frac{1}{2}
\begin{pmatrix}
\gamma_{\uparrow,\alpha}^{\dagger} &
\gamma_{\downarrow,\alpha}^{\dagger}
\end{pmatrix}
\boldsymbol{\sigma}
\begin{pmatrix}
\gamma_{\uparrow,\alpha} \\
\gamma_{\downarrow,\alpha}
\end{pmatrix} \qquad
S_{\alpha}^{+}=\gamma_{\up,\alpha}^{\dagger}\gamma_{\down,\alpha},
\qquad
S_{\alpha}^{-}=\gamma_{\down,\alpha}^{\dagger}\gamma_{\up,\alpha},
\end{equation}
so that $\ket{\alpha,\alpha}=S_{\alpha}^{-}\ket{\Uparrow}$. $H_{\textrm{eff}}$ is then the projection of a  $SU(2)$-symmetric  Heisenberg Hamiltonian given by
\begin{equation}
H_{\mathrm{S}}=\sum_{\alpha<\beta}J_{\alpha\beta}\left(\mathbf S_{\alpha}\cdot\mathbf S_{\beta}-\frac{1}{4}\right),
\label{eq:magnon_spin_model_SM}
\end{equation}
in the one magnon subspace. In particular, its one-magnon matrix elements are
\begin{equation}
\bra{\alpha,\alpha}H_{\mathrm{S}}\ket{\beta,\beta}=\frac{J_{\alpha\beta}}{2},
\qquad \alpha\neq\beta,
\end{equation}
and
\begin{equation}
\bra{\alpha,\alpha}H_{\mathrm{S}}\ket{\alpha,\alpha}=-\frac{1}{2}\sum_{\beta\neq\alpha}J_{\alpha\beta},
\end{equation}
in agreement with the exchange coupling identification in Eq.$\,$\ref{eq:J_ab_Heff}.

\begin{figure}[h!]
    \centering
    \includegraphics[width=\columnwidth]{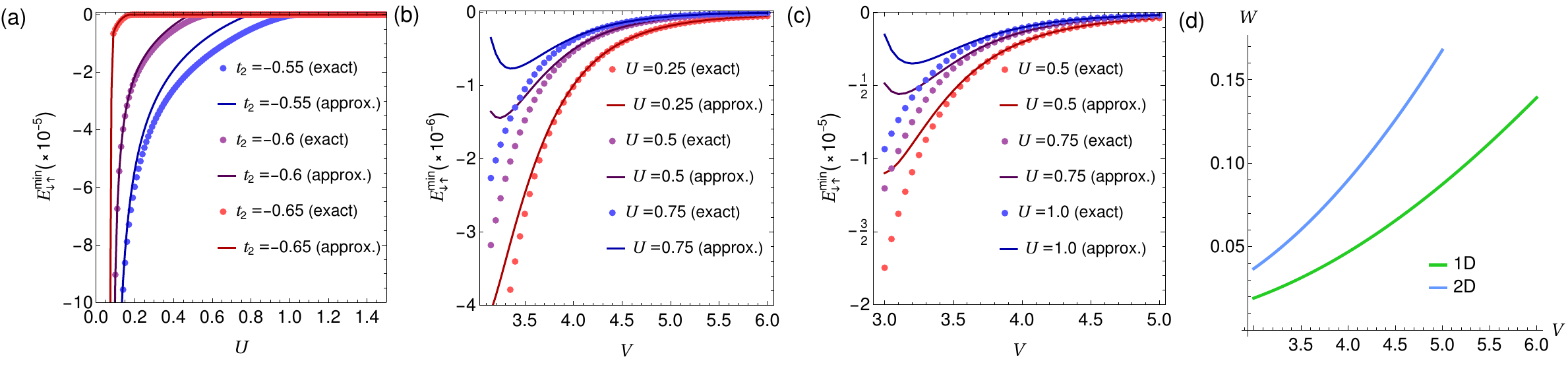}
    \caption{Comparison between the ground-state energies, $E_{\down\up}^{\textrm{min}}$, of the full spin-flip Hamiltonian in Eq.$\,$\ref{eq:magnon_H_SM_exact} and of the effective Hamiltonian in Eq.$\,$\ref{eq:magnon_spin_model_SM}. (a) Results for the 1D Aubry-André model with $V=3.5,t_3=0$ and different $t_2$, for $L=139$. The ground-state energy, as well as the interaction strength above which the ground-state becomes fully polarized (i.e. when $E_{\down\up}^{\textrm{min}}=0$) are well captured. (b)  Variable-$V$ results for the 1D Aubry-André model with $t_2=t_3=0$, for different $U$, with $L=139$.  (c)  Variable-$V$ results for the 2D Aubry-André model with $t_2=t_3=0$, for different $U$, with $L=33$. (d) Narrow-band bandwidth for the 1D and 2D Aubry-André models, for $t_2=t_3=0$ (shown for direct comparison with the magnon energy scales in the remaining plots). }
    \label{fig:magnon_gap_exact_vs_approx}
\end{figure}

We finish this section by  comparing the ground-state energy of the full two-particle Hamiltonian in Eq.$\,$\ref{eq:magnon_H_SM_exact} and of the effective Hamiltonian in Eq.$\,$\ref{eq:magnon_spin_model_SM}, in Fig.$\,$\ref{fig:magnon_gap_exact_vs_approx}. We can see that  the results become asymptotically exact as $V$ increases, in both 1D and 2D.

\pagebreak

\section{Robustness of narrow-band projection}

In the main text, we focused on studying the stability of the fully-polarized ferromagnetic state against charge and spin-flip excitations in the limit where the Hamiltonian was projected into the narrow-band. In this section, we numerically verify the stability of these results. In Sec.$\,$\ref{sec:higher_energy_projection} we start by verifying the robustness of the critical interaction strength $U_c$  against including higher-energy  single-particle states (beyond the narrow-band) on our projected Hamiltonian in Eq.$\,$\ref{eq:H_proj}. Then in Sec.$\,$\ref{sec:variable_Sz}, we will verify the robustness of the fully polarized ferromagnet beyond the simplest single magnon excitation. In particular, we will perform exact diagonalization for small system sizes in all the different possible spin sectors and provide numerical evidence for the global stability of the fully-polarized ferromagnet.

\subsection{Exact diagonalization results including higher-energy states}
\label{sec:higher_energy_projection}

To test the robustness of the narrow-band projection, we enlarge the projected Hilbert space by retaining the lowest $N_{\mathrm{states}}$ single-particle orbitals of the non-interacting problem, while keeping the total number of electrons fixed to the half-filled narrow-band value $N_{e}=N_{B}$. In this way, the fully polarized reference state still has $S_{z}^{\mathrm{max}}=N_{B}/2$, but electrons are allowed to visit higher-energy orbitals above the narrow band. We then compute using exact diagonalization the lowest many-body energy in the sector $S_{z}=S_{z}^{\mathrm{max}}-1$ and compare it with the energy $E_{\Uparrow}$ of the fully polarized state.

\begin{figure}[h!]
    \centering
    \includegraphics[width=\columnwidth]{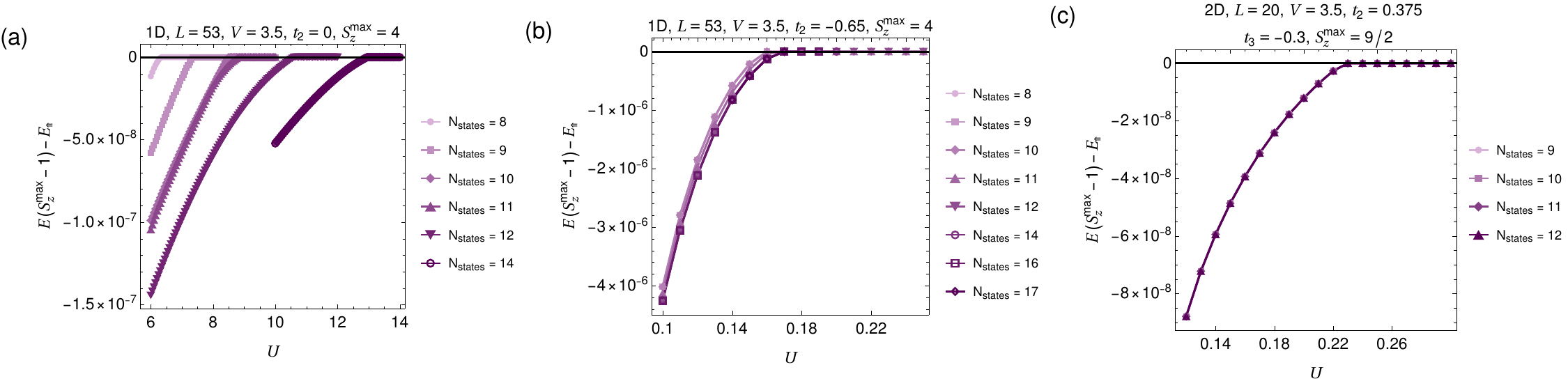}
    \caption{Robustness of the ferromagnetic transition upon enlarging the projected Hilbert space beyond the narrow-band. We plot the lowest many-body energy in the sector $S_{z}=S_{z}^{\mathrm{max}}-1$, measured relative to the fully polarized state, as a function of $U$, while retaining the lowest $N_{\mathrm{states}}$ single-particle orbitals in the projection and keeping the total electron number fixed to the half-filled narrow-band value. (a) 1D Aubry-Andr\'e model with $L=53$ (corresponding to $N_B=8$ states in the narrow-band), $V=3.5$, and $t_{2}=t_3=0$, for which the critical interaction increases strongly as higher-energy states are included. (b) 1D model with $L=53$, $V=3.5$, $t_{2}=-0.65$ and $t_3=0$, where the results are very weakly dependent on $N_{\mathrm{states}}$. (c) 2D model with $L=20$ (corresponding to $N_B=9$ states in the narrow-band), $V=3.5$, $t_{2}=0.375$, and $t_{3}=-0.3$, where the results are again essentially independent of $N_{\mathrm{states}}$. }
    \label{fig:varbmax}
\end{figure}

The results are shown in Fig.$\,$\ref{fig:varbmax}. The critical interaction $U_c$ corresponds to the interaction above which we have $E(S_{z}^{\mathrm{max}}-1)-E_{\Uparrow}=0$, i.e., the ground-state in the $S_z=S_{z}^{\mathrm{max}}-1$ spin sector becomes the $S=S_{z}^{\mathrm{max}}$ fully-polarized state. Panel Fig.$\,$\ref{fig:varbmax}(a) illustrates the bipartite 1D point $t_{2}=0$, where it can be seen that $U_c$ increases strongly as higher-energy states are added to the projection. This is consistent with the discussion in the main text and the illustration in Fig.$\,$\ref{fig:1}(c): near the bipartite limit, mixing with remote states significantly disfavors a fully polarized ground state. By contrast, panels Fig.$\,$\ref{fig:varbmax}(b,c) show non-bipartite 1D and 2D examples for parameters where $U_c$ is significantly smaller than the gap to higher-energy states, $\Delta_G$ (corresponding to particular choices of $t_2$ for the parameters studied in Fig.$\,$\ref{fig:4}). In this case, the results are essentially independent of $N_{\mathrm{states}}$, indicating that the small-$U_{c}$ ferromagnetic transition found with the narrow-band projection is well converged. These results support the picture emphasized in the main text: once the wavefunction detuning induces a $U_{c}$ well below the gap to remote states, the low-energy ferromagnetic physics becomes insensitive to band mixing.

\subsection{Exact diagonalization results for different spin sectors}
\label{sec:variable_Sz}

We next numerically verify that the local stability of the fully-polarized ferromagnet to one-spin-flip excitations indeed captures the onset of the fully polarized ferromagnet as the global ground-state (i.e. for all total spin $S$). To do so, we diagonalize the projected Hamiltonian in all the possible spin sectors with $S_{z}<S_{z}^{\mathrm{max}}$ for a fixed system size and compare the corresponding lowest energies with $E_{\Uparrow}$. 

\begin{figure}[h!]
    \centering
    \includegraphics[width=0.75\columnwidth]{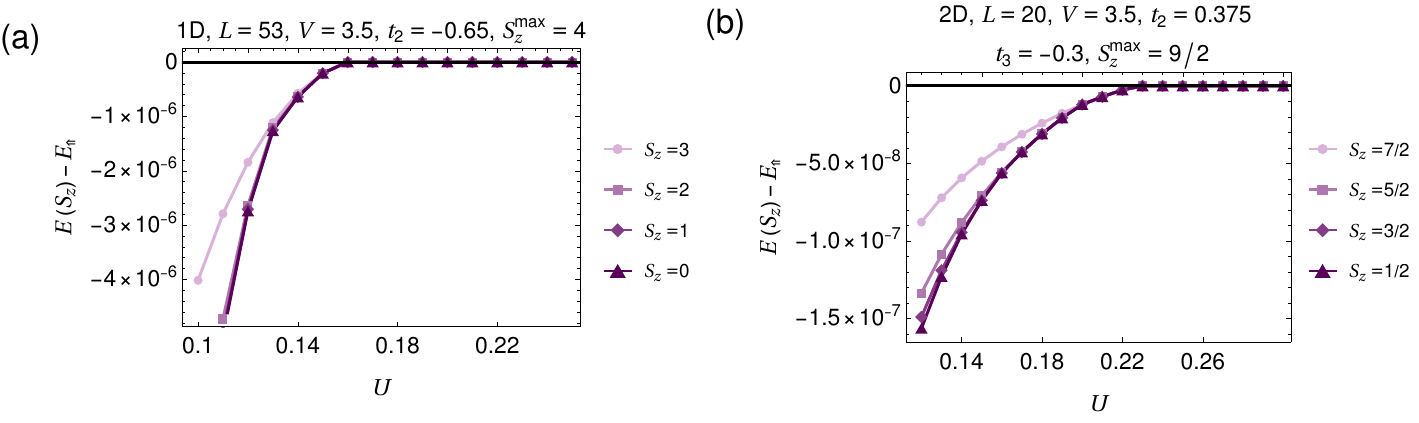}
    \caption{Lowest many-body energies in different $S_z$ spin sectors, measured relative to the fully polarized state, for the same non-bipartite parameter sets used in Fig.$\,$\ref{fig:varbmax}(b,c), retaining only the narrow-band in the projection. (a) 1D Aubry-Andr\'e model with $L=53$ ($N_B=8$), $V=3.5$, and $t_{2}=-0.65$. The plotted sectors are $S_{z}=3,2,1,0$, with $S_{z}^{\mathrm{max}}=4$. (b) 2D model with $L=20$  ($N_B=9$), $V=3.5$, $t_{2}=0.375$ and $t_{3}=-0.3$. The plotted sectors are $S_{z}=7/2,5/2,3/2,1/2$, with $S_{z}^{\mathrm{max}}=9/2$. }
    \label{fig:varSz}
\end{figure}

Figure$\,$\ref{fig:varSz} shows results for representative 1D and 2D examples. In both panels, the ground-state for all $S_{z}$ sectors above $U_c$ is the fully polarized state with $S=S_{z}^{\mathrm{max}}$ --- diagnosed by $E(S_{z})-E_{\Uparrow}=0$. At smaller fixed $U$, the energy decreases with $S_z$ and the ground-state becomes fully unpolarized. These results provide direct numerical support that the stability to one-spin-flip excitations that we focused on in the main text captures well the transition into a global fully-polarized ground-state.  

\end{document}